\documentclass[reprint,aps,pre]{revtex4-2}
\usepackage{graphicx}
\usepackage{amssymb}
\usepackage{amsmath}
\usepackage{gensymb}
\usepackage{xcolor}
\usepackage{hyperref}
\usepackage[utf8]{inputenc}
\usepackage{comment}
\usepackage{lipsum}
\usepackage[normalem]{ulem}
\usepackage{orcidlink}
\usepackage{subfigure}
\usepackage[section]{placeins}
\usepackage[margin=1.5cm]{geometry}
\begin{document}
\title{Binding of holes and competing spin-charge order in simple and extended Hubbard model on cylindrical lattice: An exact diagonalization study}
\author{Md Fahad Equbal \orcidlink{0009-0004-0054-9068}}
 \email{md179654@st.jmi.ac.in}
 \affiliation{Department of Physics, Jamia Millia Islamia (Central University), New Delhi $110025$, India}
\author{M. A. H. Ahsan \orcidlink{0000-0002-9870-2769}}
 \email{mahsan@jmi.ac.in}
\affiliation{Department of Physics, Jamia Millia Islamia (Central University), New Delhi $110025$, India}
\date{\today}
\begin{abstract}
We investigate the binding of holes and the emergence of competing spin-charge order in the simple and extended Hubbard model using exact diagonalization on the $3\times 4$ cylindrical lattice. For the simple Hubbard model ($V=0$), we find weakly bound hole pairing mediated by magnetic correlations at intermediate repulsive $U$, without any evidence of phase separation. Introducing nearest-neighbor interaction $V$ reveals a rich phase diagram: attractive $V$ drives multi-hole clustering and phase separation with localized magnetic quenching, while repulsive $V$ stabilizes charge-density-wave (CDW) order that coexists with bound hole pairs within a modulated magnetic background. At strong coupling ($U=10$), the competition sharpens, with attractive $V$ overcoming on-site repulsion to form magnetically quenched clusters and repulsive $V$ producing robust CDW order that constrains pairing. Real-space analysis of spin and charge correlations provides microscopic evidence of distinct binding mechanisms---phase separation versus correlation-mediated pairing---depending on the sign and strength of intersite interaction $V$. Our results establish a comprehensive picture of how nonlocal Coulomb interactions reshape the landscape of hole-binding and collective order in correlated electron systems.

\textbf{Keywords:} Binding energy; Superconductivity; Phase separation; Hubbard model; Exact diagonalization
\end{abstract}
\maketitle

\section{\label{intro}Introduction}
The microscopic mechanism of high-temperature superconductivity in correlated electron systems remains one of the most enduring challenges in condensed matter physics. Since the initial discovery in cuprates \cite{Bednorz1986}, unconventional superconductivity has been observed in diverse materials including heavy fermion compounds \cite{Bwhite2015}, magic-angle graphene \cite{Ycao2018}, cobaltates \cite{Kuang2013} and iron pnictides \cite{Hwen2011}, where pairing is believed to emerge from electronic correlations and magnetic fluctuations \cite{Scalapino1999,Palee2006,Davis2013,Kiemer2015} rather than conventional phonon mediation \cite{Bardeen1957}. The Hubbard model \cite{Hubbard1963,Gutzwiller1963,Kanamori1963} and its extensions serve as fundamental theoretical framework for capturing the strong correlation effects, yet despite decades of research \cite{Ying2014, Xtgrover2021, Hakivelson2021, Gzsheng2021, Christos2023, Gangsu2024, Fahad2025a, Weifeng2008}, fundamental questions persist about whether these minimal models can give rise to robust superconducting phases with purely repulsive interactions.

Recent advances have revitalized this discussion, with large-scale numerical studies and quantum simulations revealing new insights into the interplay between magnetism, charge order, and pairing. Density matrix renormalization group (DMRG) and tensor network studies \cite{Zcweng2022, Fatih2024} have identified strong two-hole binding and competing spin and charge density wave tendencies, while experimental work with ultracold atoms in optical lattices \cite{Hirthe2023, Shuzheng2024} has directly demonstrated hole pairing mediated by magnetic correlations. These developments highlight that even after extensive investigation \cite{Christos2023}, the doped Hubbard model continues to pose unresolved questions about the nature of its ground state and the microscopic origin of pairing correlations.

In this context, the formation of bound states between doped holes serves as a crucial test for any proposed pairing mechanism. A negative hole-binding energy signals effective attraction of purely electronic origin and is often viewed as a precursor to superconductivity. Early numerical studies using exact diagonalization (ED) and quantum Monte Carlo (QMC) \cite{Ggubernatis1998, Jig2021, Shengtao2023} reported negative binding energies in specific parameter regimes, suggesting that two holes can indeed bind even in purely repulsive Hubbard model. However, the stability and character of these bound states depend sensitively on both interaction parameters and lattice geometry.

The inclusion of nonlocal Coulomb interactions introduces additional complexity, with the extended Hubbard model capturing the rich interplay between on-site and intersite interactions resulting in competing orders such as charge density wave (CDW) and spin density wave (SDW) states \cite{Fahad2025b, Fahad2025c, Rmicnas1990, Schular2013}. Recent studies have shown that extended interactions qualitatively alter binding behavior \cite{Schular2013, Klemeshko2015}, while geometry plays an essential role in modifying the competition between kinetic and potential energies \cite{Dagotto1994, Scalapino2012}. Functional renormalization group (FRG) and dynamical cluster approximation (DCA) analyses \cite{Leblanc2015, Lieb1968} have emphasized how finite-size geometries connect one- and two-dimensional behaviors relevant for cuprates, yet a systematic understanding of how lattice geometry and nearest-neighbor interactions mutually affect hole-binding near phase separation remains incomplete.

This work addresses this gap through a comprehensive ED study of the simple and extended Hubbard model on a 12-site ($3\times 4$) lattice with periodic boundary conditions (PBC) along x-direction and open boundary conditions (OBC) along y-direction resulting in cylindrical geometry. We demonstrate that geometry and extended interactions crucially influence binding phenomena and the emergence of competing orders. In the simple Hubbard model, we find stable two-hole binding coexisting with SDW order for repulsive $U$, while higher-order clusters remain unbound. Introducing nearest-neighbor interaction $V$ dramatically reshapes this landscape: attractive $V$ drives multi-hole clustering and phase separation with localized magnetic quenching, while repulsive $V$ stabilizes CDW order coexisting with bound hole pairs within a modulated magnetic background. At strong coupling ($U=10$), the competition sharpens, with attractive $V$ overcoming on-site repulsion to form magnetically quenched clusters and repulsive $V$ producing robust CDW order that constrains pairing.

By systematically varying both $U$ and $V$ across attractive and repulsive regimes, and by analyzing corresponding spin and charge correlations through real-space visualization, we establish a coherent microscopic picture of how nonlocal interactions govern whether hole-binding proceeds via phase separation or occurs within charge-ordered backgrounds. These insights clarify the fundamental interplay between spin and charge channels leading to distinct pairing and ordering tendencies in minimal correlated-electron models, providing a crucial bridge between exact results from finite-size systems and the emergent behavior of extended systems. 

The remainder of this paper is organized as follows: In Section \ref{modeldesr}, we describe the Hubbard model and outline the theoretical framework to study the binding of holes leading to superconductivity. Section \ref{results} details our findings, first establishing the behavior of the simple Hubbard model ($V=0$), and then exploring how nearest-neighbor interactions ($V$) reshape the landscape of hole-binding and competing spin-charge order at intermediate ($U=4$) and strong ($U=10$) coupling. Finally, Section \ref{summary} provides a comprehensive summary of our findings.
\section{\label{modeldesr}Model and Method}
The extended Hubbard model is considered the workhorse to study strongly correlated electron systems in modern condensed matter physics. It is the simplest model which takes into account the competition between the kinetic enrgy and the potential energy corresponding to the on-site and the intersite interactions. The Hamiltonian for the one-band extended Hubbard model on a 2D lattice in real space is defined as \cite{Rmicnas1990}

\begin{equation}
H = -t \sum_{\langle ij\rangle,\sigma} (c_{i\sigma}^\dagger c_{j\sigma} + h.c.)
  + U \sum_i n_{i\uparrow} n_{i\downarrow}
  + V \sum_{\langle ij\rangle} n_i n_j ,
\label{hamil}
\end{equation}
where $c_{i\sigma}^\dagger (c_{i\sigma})$ is the fermionic operator that creates (annihilates) an electron with spin $\sigma \in \{\uparrow, \downarrow\}$ at lattice site $i$, and $\langle ij\rangle$ denotes nearest-neighbor (NN) sites on the lattice. The parameters $t$, $U$ and $V$ represents the NN hopping amplitude, the on-site (local) Coulomb interaction and the intersite (nonlocal) Coulomb interaction, respectively. The number operator $n_{i\sigma} = c_{i\sigma}^\dagger c_{i\sigma}$ counts particles at site $i$ with spin $\sigma$, $n_i = n_{i\uparrow} + n_{i\downarrow}$ and $h.c.$ means Hermitian conjugate. In the $V=0$ limit, the Hamiltonian $H$ in Eq. (\ref{hamil}) reduces to the simple Hubbard model. We denote the number of lattice sites by $M$, electrons by $N_e$ and holes by $N_h$ with filling factor defined as $n=N_e/M$. The half-filled system corresponds to $n=1$. The dispersion relation for an infinite system corresponding to the non-interacting ($U=V=0$) case in 2D is given by

\begin{equation}
\epsilon_k = -2t(cosk_x + cosk_y),
\label{dispersion}
\end{equation}
\noindent 
with bandwidth $W=8t$.

Various analytical and numerical techniques have been employed to study the Hubbard model. Despite its simplistic appearance, the model can only be solved exactly in one dimension (1D) where the Bethe ansatz offers an exact solution \cite{Lieb1968, Lieb2003}, though it still leaves some aspects of the system unclear. The lack of an analytic solution to the Hubbard model in two or higher dimensions has resulted in the development of various numerical methods including mean-field theory (MFT) \cite{Penn1966, Yann2014}, cellular dynamical mean-field theory (CDMFT) \cite{Cwalsh2023}, density matrix renormalization group (DMRG) techniques \cite{Shengtao2023},  quantum Monte Carlo (QMC) simulations \cite{Varney2009, Shuhui2020}, and variational approaches \cite{Gutzwiller1965} to study its low-temperature and ground state properties. One particularly effective numerical method for solving the Hubbard model is exact diagonalization (ED). This method accurately computes the ground state and the low-lying excited states but is only feasible for small clusters due to the exponential growth of the Hilbert space. We utilize spin-adapted basis \cite{Sarma1996} to calculate the ground state and low-lying states which significantly reduces the dimensionality of the Hilbert space. 

The idea of hole-binding is particularly significant in the context of high-temperature superconductivity, where certain theories suggest that electron pairs (Cooper pairs) that are responsible for superconductivity can arise through the binding of holes in a strongly correlated electron system such as the Hubbard model. To investigate the possibility of superconductivity in the system, we study the binding energy of a pair of hole defined as \cite{Jariera1989, Edagottom1990}
\begin{equation}
E_{B2} = (E(2)-E(0))-2(E(1)-E(0)),
\label{BE2}
\end{equation}
\noindent
where $E(0)$ is the ground state energy for the half-filled case and $E(1)$, $E(2)$ are the ground state energies with one and two holes respectively. The bound state of two holes which may lead to superconductivity is indicated by a negative value of $E_{B2}$. Equation (\ref{BE2}) signifies that if binding occurs then the ground state energy of two-hole doped together into a half-filled band will be lower than if the two holes were doped separately to a half-filled system. However, it has been conjectured that the region of parameter space in which superconductivity is most likely to occur is in the vicinity of phase separation \cite{Emkivelson1993, EDagotto1994}. We therefore, also study the binding energy of three and four holes \cite{Jariera1989, Callaway1990}
\begin{equation}
E_{B3} = (E(3)-E(0))-(E(2)-E(0))-(E(1)-E(0)),
\label{BE3}
\end{equation}
\begin{equation}
E_{B4} = (E(4)-E(0))-2(E(2)-E(0)),
\label{BE4}
\end{equation}
where $E(3)$ and $E(4)$ are the ground state energies with three and four holes respectively. If both $E_{B3}$ $<$ 0 and $E_{B4}$ $<$ 0, phase separation occurs. Thus, superconductivity is characterized by $E_{B2}$ $<$ 0 and $E_{B3}$, $E_{B4}$ $>$ 0. We observe that the inclusion of first neighbor interaction $V$ in the Hubbard model significantly affects binding, alongside the manifestation of phase separation.
\section{\label{results}Results and Discussion}
To elucidate the emergence of pairing and phase separation in correlated electron systems, we systematically investigate the energetics and correlation properties of the simple and extended Hubbard models on the $3\times 4$ cylindrical lattice with hole dopings $N_h=0-4$ using exact diagonalization. Our analysis proceeds by first examining the simple Hubbard model ($V=0$), which serves as a reference for identifying the intrinsic mechanism of hole-binding and the evolution of spin-charge correlations with interaction strength. We then extend the study to include a finite nearest-neighbor Coulomb interaction $V$, allowing us to explore how it reshape the balance between local pairing, charge order and magnetic correlations.

\subsection{Simple Hubbard model}

\begin{figure}[h]
\centering
\includegraphics[scale=0.57]{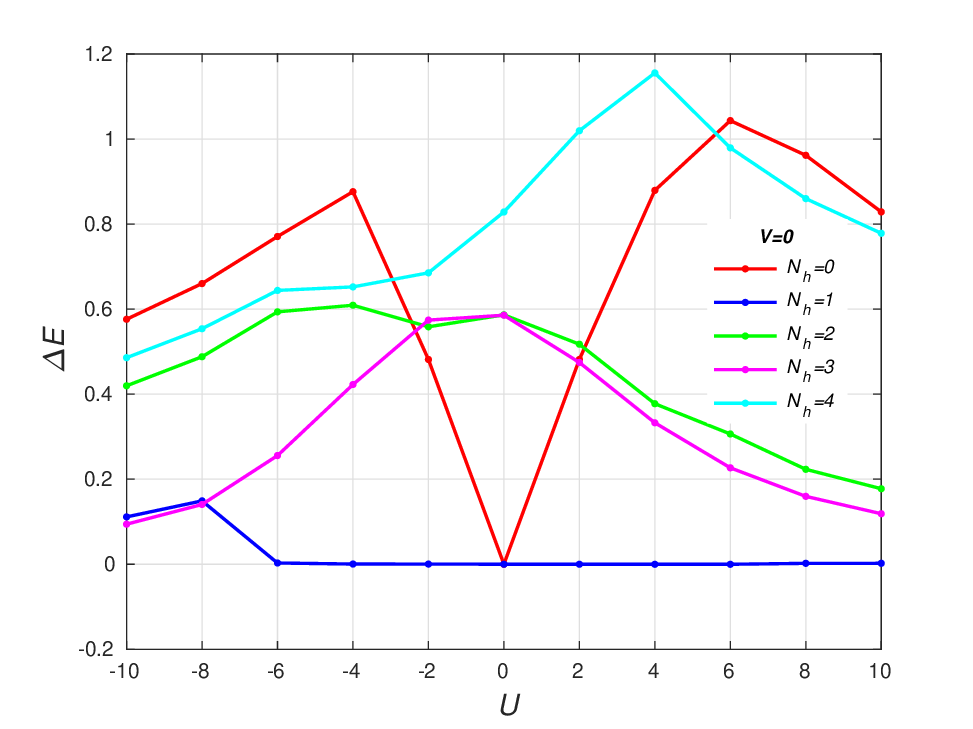}
\caption{The excitation gap $\Delta E$ as a function of $U$ for the half-filled ($N_h=0$), one-hole ($N_h=1$), two-hole ($N_h=2$), three-hole ($N_h=3$) and four-hole ($N_h=4$) doped systems of the $3\times 4$ cylindrical lattice.}
\label{fig:egappbc}
\end{figure}

Figure \ref{fig:egappbc} shows the excitation gap $\Delta E=E_1 - E_0$, where $E_1$($E_0$) is the first excited(ground) state energy, as a function of $U$ for the $3\times 4$ cylindrical lattice at hole dopings $N_h=0-4$. At half-filling ($N_h=0$), $\Delta E$ vanishes at $U=0$ and increases on both the repulsive and attractive sides before decreasing again at intermediate $|U|$. For the one-hole doped system ($N_h=1$), $\Delta E$ remains nearly zero across most of the $U$ range but acquires a finite value for $U< -6$, reflecting the onset of local pairing and charge localization in the strongly attractive regime, where pair-formation suppresses spin fluctuations and produces a gapped spin-charge background. The two-hole ($N_h=2$) and the three-hole ($N_h=3$) doped systems exhibit an intermediate value of $\Delta E$ in the weak-coupling regime ($-2\lesssim U\lesssim 2$) that decrease steadily in both strongly attractive and strongly repulsive limits. The four-hole ($N_h=4$) doped system shows the largest $\Delta E$ at $U=4$, beyond which the gap decreases symmetrically as $U$ moves toward either the strongly repulsive or weak-coupling to attractive regimes. The pronounced peak of $\Delta E$ at $U=4$ marks the crossover to an intermediate-coupling regime in which spin correlations are maximized while charge motion is restricted by on-site repulsion $U$. Beyond this point, the gap decreases as spin exchange weakens ($J\propto 4t^2/U$), signaling the transition toward a localized Mott-like state with reduced spin correlations. These trends underscore the competing nature of spin and charge excitations across interaction strengths and doping.

\begin{figure}[h]
\centering
\includegraphics[scale=0.57]{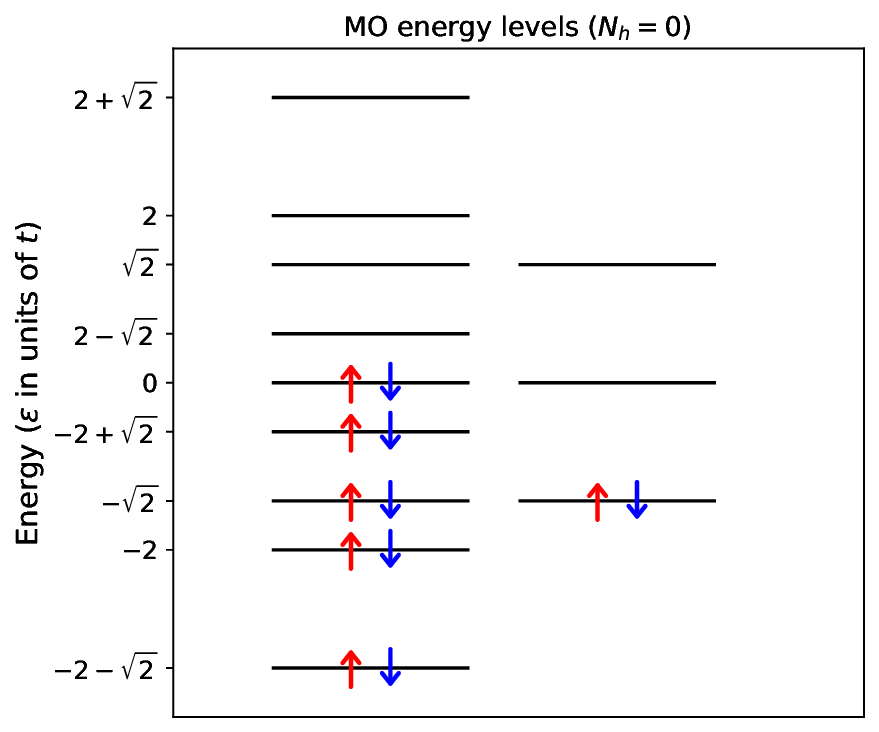}
\caption{Single-particle molecular orbital (MO) energy spectra of the non-interacting ($U=0$) half-filled ($N_h=0$) Hubbard model on the $3\times 4$ cylindrical lattice. Degeneracy at the Fermi level for the $N_h=0$ and $N_h=1$ cases yields a vanishing gap.}
\label{fig:molevels}
\end{figure}    

To understand the distinct behavior observed in the excitation gap across hole dopings $N_h=0-4$, we analyze the non-interacting ($U=0$) molecular-orbital (MO) energy levels shown in Fig. \ref{fig:molevels}. At half-filling, the Fermi level lies within the doubly degenerate orbitals at $\epsilon=0$, giving rise to degenerate highest occupied and lowest unoccupied states; thus the excitation gap $\Delta E$ vanishes at $U=0$ and opens up slightly with finite $U$ as the degeneracy is lifted. Removing one electron from $\epsilon =0$ leaves a single hole in this degenerate manifold, resulting in a quasidegenerate set of low-energy configurations and maintaining a nearly zero gap across a range of $U$ values. When two electrons are removed from $\epsilon=0$, a closed shell forms with the highest occupied level at $\epsilon=-2+\sqrt{2}$; the resulting single-particle separation $2-\sqrt{2}$ gives rise to a large excitation gap, consistent with the robust $\Delta E$ observed for the two-hole system. For three-hole system, one electron occupies the non-degenerate $\epsilon=-2+\sqrt{2}$, leading to an intermediate value of excitation gap. For the four-hole system, the occupation extends only up to the doubly degenerate $\epsilon=-\sqrt{2}$, which is again a closed-shell configuration yielding the largest $\Delta E$ in the intermediate $U$ regime. This sequence of open- and closed-shell fillings at $U=0$ explains the excitation gap behaviour and shows that the correlation-driven spectrum at finite $U$ inherits its structure from the underlying MO hierarchy. The near-zero one-hole gap together with the well-separated two-hole ground state is a characteristic finite-cluster signature of distinct one- and two-hole low-energy physics and motivates the binding-energy analysis that follows.

\begin{figure}[h]
\centering
\includegraphics[scale=0.64]{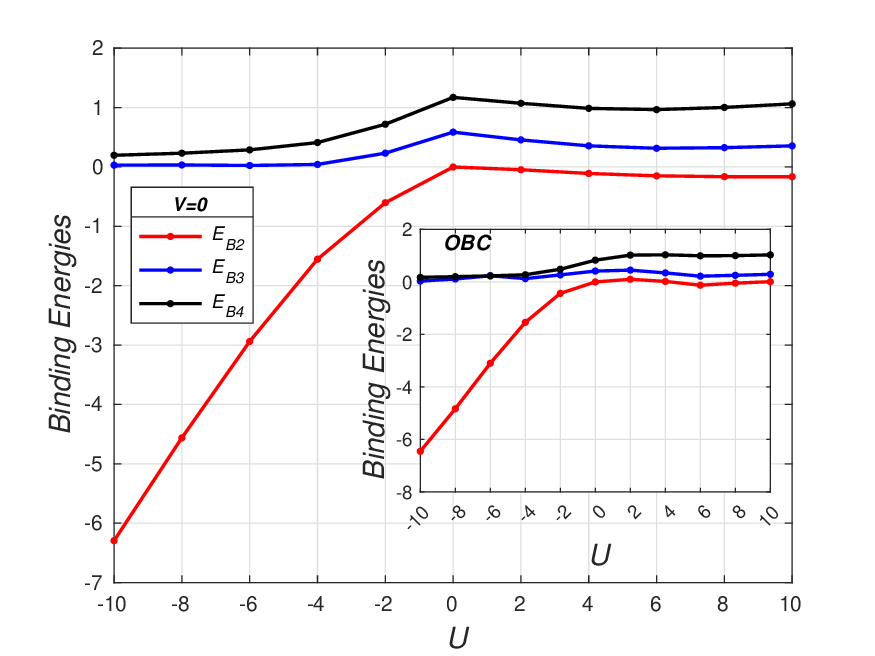}
\caption{Two-, three- and four-hole binding energies as a function of $U$ in the ground state of the Hubbard model with $N_h=2$, 3 and 4 holes, respectively, on $3\times 4$ cylindrical lattice. The inset compares these results with the corresponding lattice with open boundary conditions (OBC).}
\label{fig:behs}
\end{figure}

Figure \ref{fig:behs} shows the two-, three- and four-hole binding energies ($E_{B2}$, $E_{B3}$ and $E_{B4}$ respectively), plotted as a function of $U$. A negative binding energies indicate an effective attraction between doped holes and a tendency toward pairing or phase separation. At $U=0$, the binding energies are $E_{B2}=0$, $E_{B3}=2-\sqrt{2}$ and $E_{B4}=2(2-\sqrt{2})$. For $U<0$, the two-hole binding is large and negative (e.g., $E_{B2}\approx -6.3$ at $U=-10$), indicating strong tendency for two-hole pairing, driven by the attractive interaction. In the repulsive regime ($U>0$), $E_{B2}$ becomes only weakly negative ($\sim$-0.11 at $U=4$, and $\sim$-0.16 at $U=8-10$), indicating a modest correlation-induced pairing tendency rather than a strong, tightly bound pair. The $E_{B3}$ and $E_{B4}$ are positive for all $U$. Thus, we find that the $3\times 4$ cylindrical lattice prefers the formation of hole-pairs rather than trimers or quartets, and does not show unambiguous phase separation on this finite-size lattice. We note, however, that the magnitudes of $E_{B2}$ in the repulsive regime are small and sensitive to boundary conditions --- indeed, the lattice with OBC shown in the inset exhibit negative $E_{B2}$ only in the range $6\leqslant U\leqslant 8$.

The pairing mechanism of holes within a lattice model can also be understood from the interplay between hole delocalization and the ordering of spin-charge states. When a hole hops between lattice sites, it displaces the spin as well as charge, resulting into an energetically less favorable magnetic structure. The displaced spin-charge configuration surrounds the single hole, creating a magnetic polaron with restricted mobility \cite{Jig2021, Koepsell2019}. When a second hole hops along the path of the first hole, it can restore the spin and charge sector to its original state. This allows the two holes to form a highly mobile bound pair. 

To probe the short-range magnetic and charge environment experienced by doped holes we compute two-point spin-spin and charge-charge correlation functions:
\begin{align}
L_{ij} &= \frac{1}{4} \langle G|(n_{i\uparrow} - n_{i\downarrow})(n_{j\uparrow} - n_{j\downarrow}) |G\rangle, \\
D_{ij} &= \langle G|n_i n_j |G\rangle,
\end{align}
where $|G\rangle$ represents the ground state vector. As the number of lattice sites increases, so does the complexity of the correlation functions between different sites. It is, therefore, useful to define the corresponding structure factors,

\begin{equation}
S_X(\mathbf{q}) = \frac{1}{M}\sum_{ij} e^{i\mathbf{q}(\mathbf{R_i} - \mathbf{R_j})} X_{ij},
\label{chargsf}
\end{equation}
\noindent
where $X\in\lbrace L,D\rbrace$, $\mathbf{q}$ is the wave vector and $\mathbf{R_i}$($\mathbf{R_j}$) denote the position of the lattice site $i$($j$). To probe correlations associated with antiferromagnetic or charge-density-wave order on the 2D square lattice, we focus on the ordering wave vector $\mathbf{q} = (\pi, \pi)$.

\begin{figure}[h]
\centering
\includegraphics[scale=0.64]{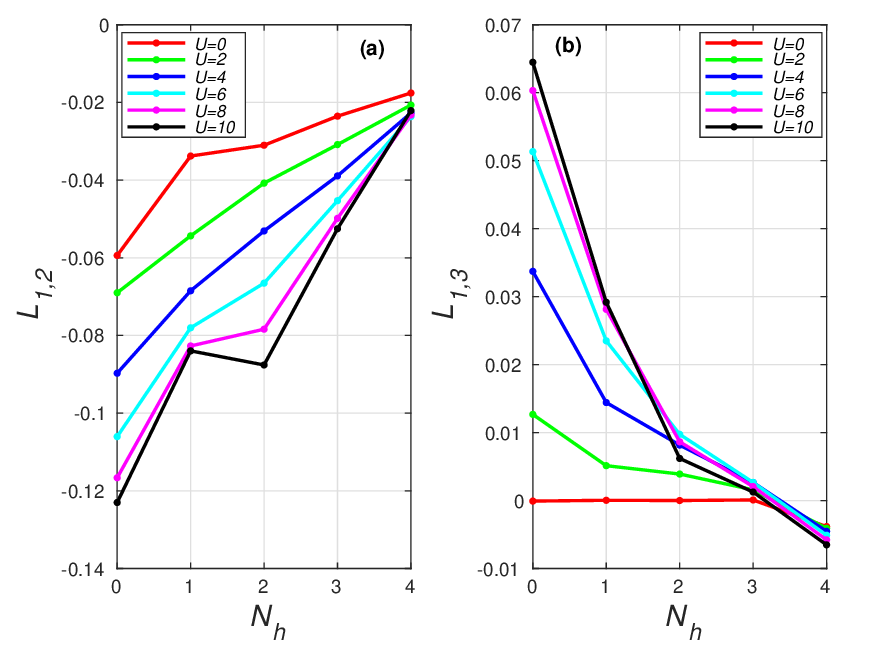}
\caption{Comparison of (a) nearest-neighbor spin correlation ($L_{1,2}$) and (b) next nearest-neighbor spin correlation ($L_{1,3}$), as a function of number of holes $N_h$ at $U=0$ and several positive $U$ values on $3\times 4$ cylindrical lattice.}
\label{fig:L1213shm}
\end{figure}

Figure \ref{fig:L1213shm} shows the nearest-neighbor ($L_{1,2}$) and next nearest-neighbor ($L_{1,3}$) spin-spin correlation functions for $N_h=0-4$ on the $3\times 4$ cylindrical lattice at different $U$ values. For the noninteracting ($U=0$) case, both $L_{1,2}$ and $L_{1,3}$ are negative for all dopings, indicating an overall antiferromagnetic alignment inherited from the single-particle states. For repulsive interactions $(U>0)$, $L_{1,2}$ remains negative and its magnitude increases with $U$, signifying the strengthening of short-range antiferromagnetic correlations as electron localization sets in. In contrast $L_{1,3}$, changes sign and becomes positive for $N_h=0-3$ beyond $U\geqslant 2$, while remaining negative for $N_h=4$. This sign reversal suggests that moderate hole doping induces local spin rearrangements---possibly short-range canting or spin texture formation---within an otherwise antiferromagnetic background. The persistence of strong negative $L_{1,2}$ together with a positive $L_{1,3}$ in the two-hole sector coincides with the most pronounced hole-binding observed in Fig. \ref{fig:behs}, implying that pairing arises within a magnetically correlated environment that accommodates the motion of bound holes. For higher dopings, both correlators weaken in magnitude, consistent with the suppression of spin correlations and the emergence of charge-dominated configurations that precede phase separation, consistent with the positive three- and four-hole binding energies. In the attractive ($U<0$) regime (results not shown) the spin correlations are nearly quenched, reflecting trivial on-site pairing rather than the correlation-driven intertwined state seen for repulsive $U$.  
 
\begin{figure}[h]
\centering
\includegraphics[scale=0.64]{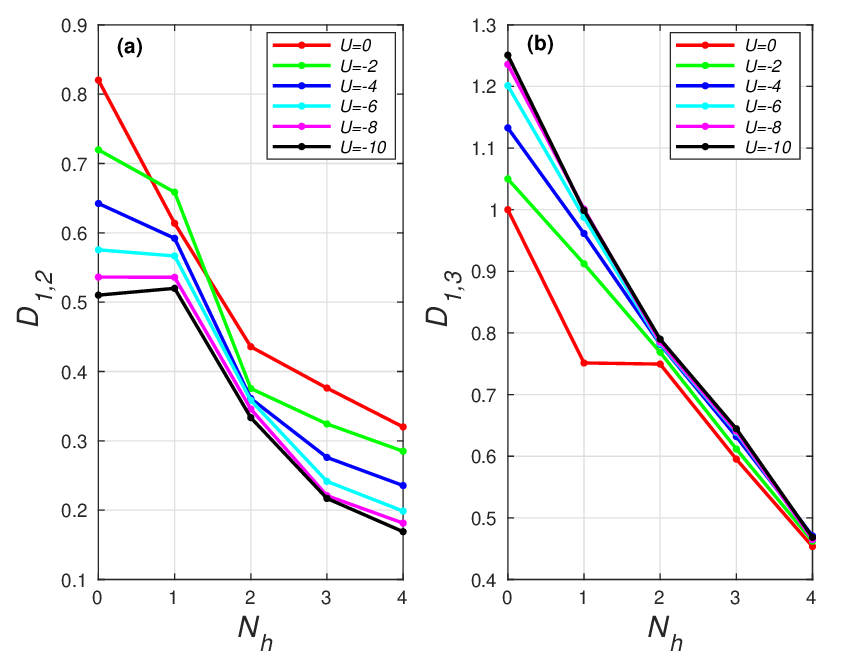}
\caption{Comparison of (a) nearest-neighbor charge correlation ($D_{1,2}$) and (b) next nearest-neighbor charge correlation ($D_{1,3}$), as a function of number of holes $N_h$ at $U=0$ and several negative $U$ values on $3\times 4$ cylindrical lattice.}
\label{fig:D1213shm}
\end{figure}  

To complement the analysis of short-range spin correlations, we examine the charge-charge correlation functions between nearest-neighbour ($D_{1,2}$) and next nearest-neighbour ($D_{1,3}$) sites. These quantities capture the degree of local charge inhomogeneity and are essential to understanding the microscopic origin of hole-binding and phase separation. Figure \ref{fig:D1213shm} presents $D_{1,2}$ and $D_{1,3}$ as functions of hole number $N_h$=0-4 for $U$=0,-2,-4,-6,-8 and -10 on the $3\times 4$ cylindrical lattice. For $U\leqslant 0$ regimes, both correlators evolve systematically with increasing $|U|$. As the attraction strengthens, $D_{1,2}$ decreases while $D_{1,3}$ increases for all the systems. This indicates a redistribution of charge at short distances consistent with the formation of tightly bound localized on-site pairs that reduce simultaneous occupation on neighbouring sites while enhancing correlations at slightly longer separations. These trends are wholly consistent with the large negative binding energies found for $U<0$ shown in Fig. \ref{fig:behs}. Turning to the repulsive regime (results not shown), the charge correlators show a different character. At half filling $D_{1,2}$ increases slowly with $U$, while $D_{1,3}$ remains nearly constant and close to unity. For doped systems $D_{1,2}$ typically grows with $U$, whereas $D_{1,3}$ shows mild decreases for several dopings. This behaviour indicates suppressed local charge fluctuations with increasing repulsion and a trend toward a more uniform (locally less clustered) charge distribution.

These charge-correlation trends dovetail with our earlier spin-correlation and binding-energy analyses. In the attractive regime the strong increase of $D_{1,3}$ together with the reduction of $D_{1,2}$ mirrors the large negative $E_{B2}$ and the near-vanishing $L_{1,2}$, $L_{1,3}$ magnitudes: holes are bound by local attraction and spin degrees of freedom are quenched. In the repulsive regime, by contrast, the binding energies are small or only weakly negative and the spin correlators $L_{1,2}$ remain appreciably negative while $L_{1,3}$ often changes sign --- a combination that points to correlation-mediated pairing. Specifically, at intermediate repulsive $U$ ($\sim 6-8$), the charge correlators are not as locally enhanced as in the negative $U$ case; instead they indicate moderate, extended charge coherence. Together with a still-negative $L_{1,2}$ (robust NN antiferromagnetism) and the sign-reversal of $L_{1,3}$, this pattern supports a picture in which holes form spatially extended, magnetically assisted pairs rather than tightly bound on-site bipolarons.

\begin{figure*}[!htbp]
\centering
\includegraphics[width=\textwidth]{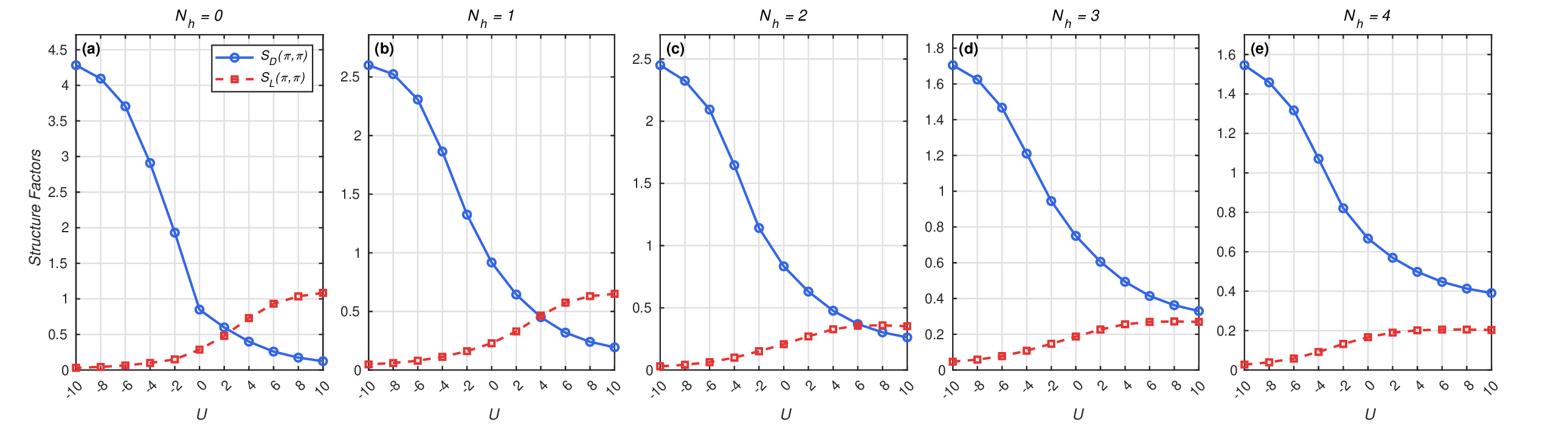}
\caption{The charge structure factor $S_D(\pi,\pi)$ and spin structure factor $S_L(\pi,\pi)$ as a function of $U$ for the $3\times 4$ cylindrical lattice. Panels (a)-(e) correspond to hole numbers $N_h$=0,1,2,3 and 4, respectively.}
\label{fig:shcssf}
\end{figure*}

To establish the collective signatures of spin and charge correlations revealed by the local observables, we compute the structure factors at the nesting vector $\mathbf{q}=(\pi,\pi)$. These quantities quantify the degree of charge-density-wave (CDW) and spin-density-wave (SDW) order and provide a bridge between the short-range correlations and the global ordering tendencies of the system. Figure \ref{fig:shcssf} presents charge structure factor $S_D(\pi,\pi)$ and spin structure factor $S_L(\pi,\pi)$ as functions of $U$ for hole dopings $N_h$=0-4. 
At half-filling ($N_h=0$), $S_D(\pi,\pi)$ decreases steadily as $U$ varies from strongly attractive ($U=-10$) to strongly repulsive ($U=10$) regimes, correspondingly $S_L(\pi,\pi)$ increases monotonically from 0.03 to 1.08. This contrasting evolution indicates a continuous suppression of charge fluctuations and a strengthening of antiferromagnetic correlations as the system evolves from a charge-ordered state in the attractive regime to a Mott-like spin-ordered state under strong repulsion. 
With finite hole doping, both structure factors are reduced in magnitude, reflecting the disruption of perfect $(\pi,\pi)$ order. For the one- and two-hole doped systems, $S_D(\pi,\pi)$ remains relatively large in the weakly attractive regime where the binding energy is most negative, signifying the dominance of charge correlations associated with hole pairing. At intermediate repulsive interactions ($U\sim 6-8$), $S_D(\pi,\pi)$ and $S_L(\pi,\pi)$ attain comparable magnitudes, consistent with the coexistence of residual spin correlations and enhanced charge fluctuations that favor bound-pair formation.
Further hole doping ($N_h$=3,4) reduces both structure factors, with $S_L(\pi,\pi)$ becoming particularly small, indicating a progressive loss of long-range spin coherence and a tendency toward charge inhomogeneity or phase separation. Overall, the evolution of $S_D(\pi,\pi)$ and $S_L(\pi,\pi)$ establishes a coherent picture with the binding energy and correlation-function results: hole-binding arises from the cooperative interplay between local spin correlations and nonlocal charge fluctuations, which dominate in different regimes of interaction strength and filling.

\begin{figure}[h]
\centering
\includegraphics[scale=0.25]{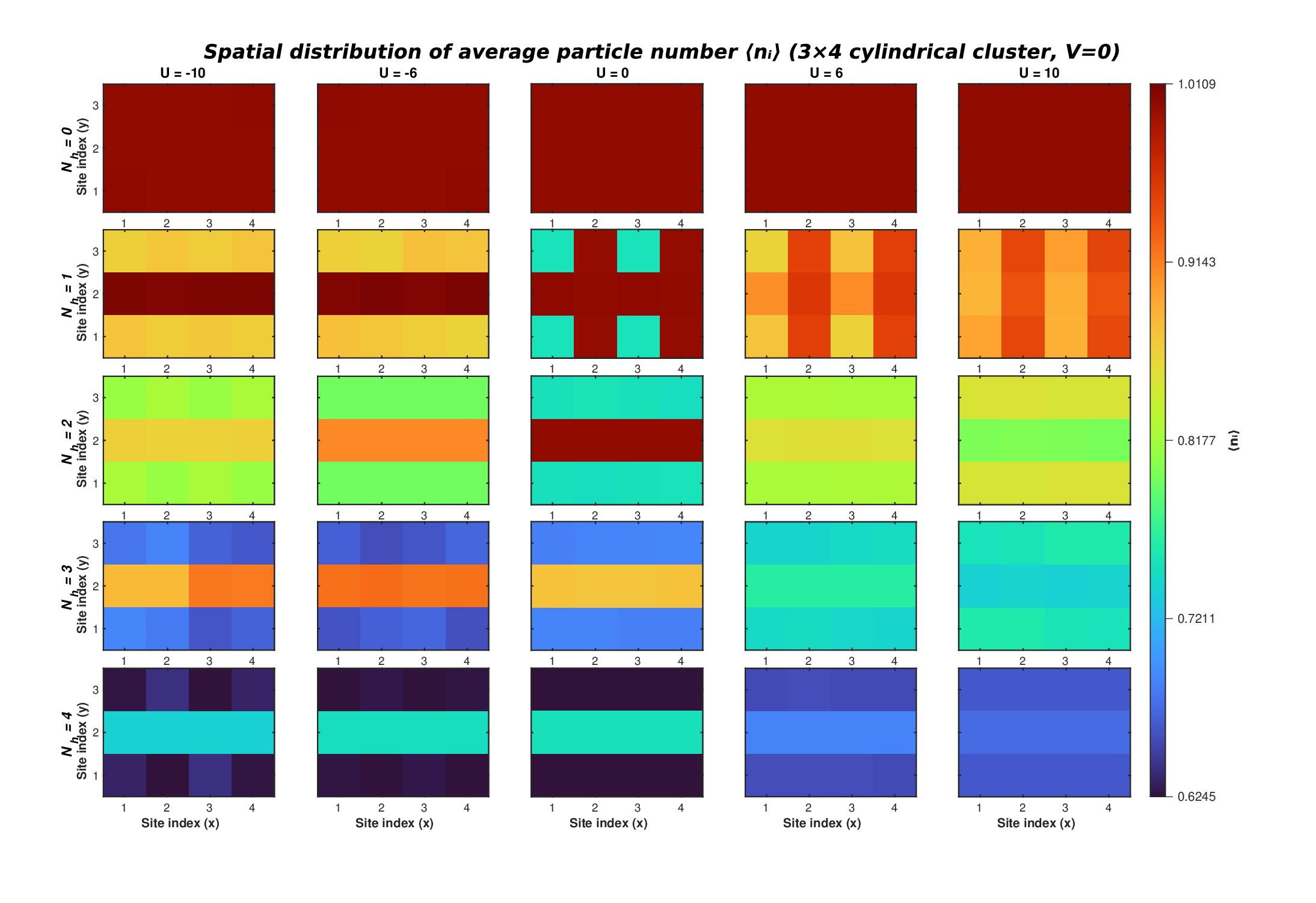}
\caption{Spatial distribution of the average site occupation $\langle n_i\rangle$ for the $3\times 4$ cylindrical lattice at interaction strengths $U$ = -10, -6, 0, 6, 10 (columns) and for hole dopings $N_h$ = 0-4 (rows). The color scale is common to all panels, ranging from 0.6245 to 1.0109. The data highlight the emergence of local charge inhomogeneities and clustering of holes in the attractive regime, contrasting with the nearly homogeneous density at large repulsive $U$.}
\label{fig:avgnishm}
\end{figure}

To gain real-space insight into the interplay between hole-binding and spin-charge ordering, we visualize the spatial pattern of the average site occupation $\langle n_i\rangle$ for the $3\times 4$ cylindrical lattice in Fig. \ref{fig:avgnishm}. At half-filling ($N_h=0$), the electron density remains nearly uniform for all $U$, as expected from the symmetric band filling. However, weak charge modulations begin to appear at large $|U|$ due to enhanced correlations. In the attractive regime ($U<0$), $\langle n_i\rangle$ shows the development of localized high-density regions even for small doping, indicating that the holes prefer to cluster and form bound states---a real-space reflection of the negative binding energies observed in the intermediate $U$ range ($|U|\approx$ 6-8). For $N_h$ = 1-2, these charge accumulations are spatially correlated with the regions where the spin correlations $L_{1,2}$ are strongest, suggesting that bound hole pairs coexist with local spin-density modulations. In contrast, in the repulsive regime ($U>0$), the charge density remains relatively homogeneous even with finite doping, consistent with the suppression of binding and the growth of antiferromagnetic correlations inferred from the spin structure factors $S_L(\pi,\pi)$. For $N_h$ = 3-4, the $\langle n_i\rangle$ maps reveal weak stripe-like modulations along the direction of open boundaries, consistent with the small but finite charge ordering captured by the charge structure factor $S_D(\pi,\pi)$. These spatial features thus unify the picture drawn from the excitation gap, binding energies, and correlation functions---showing that pairing tendencies and competing spin-charge textures emerge naturally in the intermediate $U$ domain and fade in the strong repulsive limit.

In summary, our analysis of the simple Hubbard model ($V=0$) on $3\times 4$ cylindrical lattice demonstrates a rich interplay between pairing, magnetism and charge organization. The excitation gap and binding energy analyses identify a crossover near intermediate $U$, where bound-hole states coexist with both spin- and charge-density modulations. The local spin correlations ($L_{1,2}$ and $L_{1,3}$) and charge correlations ($D_{1,2}$ and $D_{1,3}$) reveal complementary trends---pairing-driven enhancement of short-range charge correlations and concurrent suppression of long-range spin order in the attractive regime, followed by the revival of antiferromagnetic correlations for $U>0$. The real-space charge maps of $\langle n_i\rangle$ confirm that hole-binding manifests as localized charge clustering, which gradually disappears as repulsion dominates. These findings collectively establish the microscopic foundation for competing spin-charge order in the intermediate $U$ regime and provide a natural baseline for exploring the additional effects of nearest-neighbor Coulomb interactions in the extended Hubbard model, discussed next. 

\subsection{Extended Hubbard model: $U=4$}
To examine how nonlocal Coulomb interactions influence the pairing tendencies and charge organization observed in the simple Hubbard model, we first analyze the extended Hubbard model at fixed on-site repulsion $U=4$. This intermediate-coupling regime is particularly significant because, in the absence of $V$, the system already exhibits clear signatures of hole-binding and competing spin-charge correlations. By varying the nearest-neighbor interaction $V$ from -10 to 10, we can distinguish how attractive and repulsive intersite interactions, enhance or suppress these competing correlations.

\begin{figure}[h]
\centering
\includegraphics[scale=0.64]{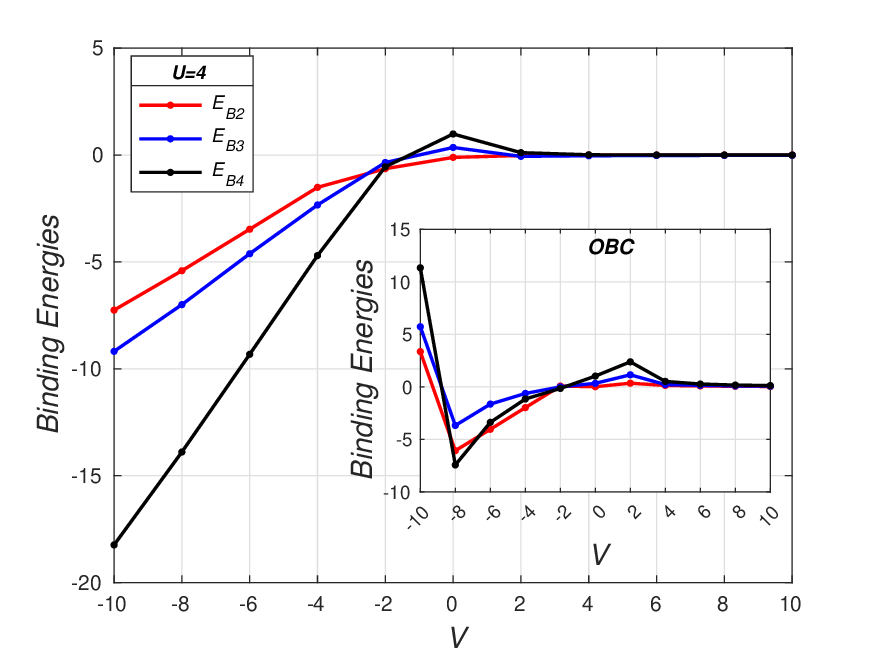}
\caption{Two-hole ($E_{B2}$), three-hole ($E_{B3}$) and four-hole ($E_{B4}$) binding energies as a function of nearest-neighbor interaction $V$ for the extended Hubbard model at $U=4$ on $3\times 4$ cylindrical lattice. The inset compares the corresponding results for the lattice with open-boundary conditions. For the cylindrical lattice, all binding energies become negative for $V\leqslant -2$, signifying strong hole clustering, whereas for the lattice with OBC, the boundary effects shift the instability to weakly attractive region.}
\label{fig:be234u4}
\end{figure}

Figure \ref{fig:be234u4} shows the two-, three- and four-hole binding energy ($E_{B2}$, $E_{B3}$ and $E_{B4}$, respectively) at fixed on-site repulsion $U=4$ as a function of the nearest-neighbor interaction $V$ for the $3\times 4$ cylindrical lattice with the inset comparing the corresponding lattice with open boundary conditions (OBC) to assess boundary effects. For the cylindrical lattice, $E_{B2}$ remains negative throughout the entire range of $V$, though its magnitude is very small in the repulsive region ($V>0$) and increases steadily with decreasing $V$, signaling enhanced two-hole binding under attractive intersite interaction. The three-hole binding energy $E_{B3}$ is negative at all $V$ values except at $V=0$, where it turns slightly positive. Its rapid increase in magnitude with more negative $V$ points to an increased tendency for hole clustering. The four-hole binding energy $E_{B4}$ behaves similarly: it is positive for $V\geqslant 0$ but turns negative for $V\leqslant -2$, with its strength increasing sharply as $V$ becomes more attractive. This collective behavior---where all binding energies become negative in the range $V\leqslant -2$---marks a crossover from a weakly correlated pairing regime to a phase-separated state driven by strong intersite attraction. In contrast, the lattice with OBC exhibits more boundary-sensitive behavior. For $V=0-10$, all three binding energies are positive, indicating the absence of stable bound-hole formation. As $V$ decreases, $E_{B4}$ becomes negative first at $V=-2$, followed by $E_{B2}$ and $E_{B3}$ at $V\leqslant -4$, where all three become simultaneously negative and their magnitudes grow rapidly, signaling phase separation consistent with the cylindrical lattice. At $V=-10$, the binding energies become positive again, suggesting that the extreme attraction drives hole localization rather than collective phase separation, likely due to finite-size and boundary effects. Overall, these trends confirm that moderate attractive $V$ promotes real-space hole pairing and cluster formation, while large repulsive $V$ suppresses it, stabilizing uniform charge configurations.

\begin{figure}[h]
\centering
\includegraphics[scale=0.285]{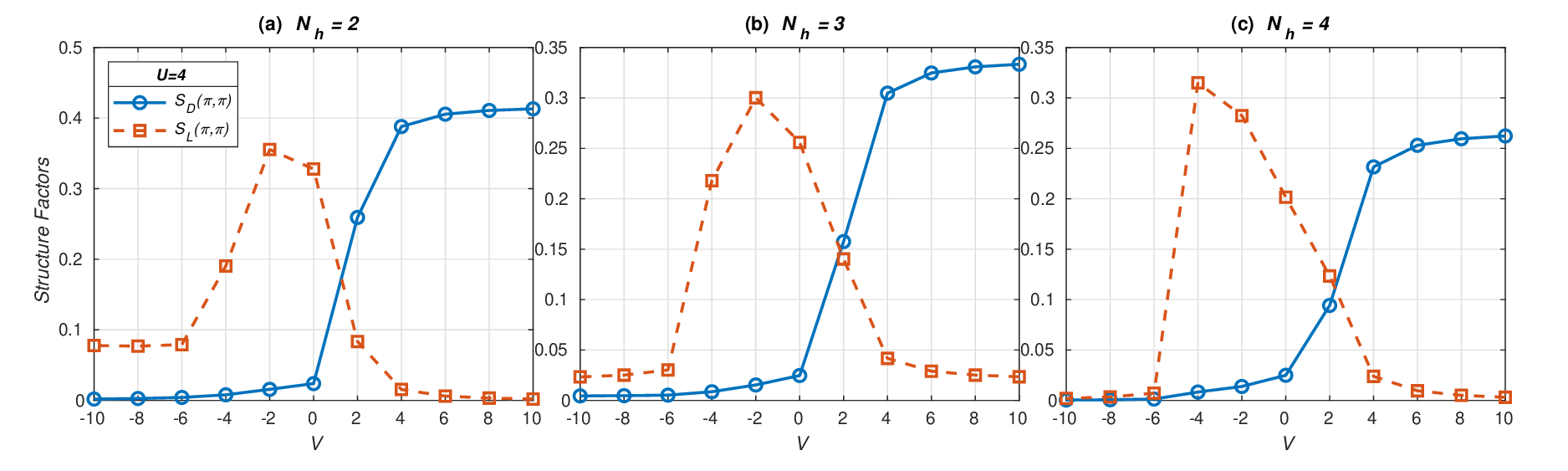}
\caption{Charge structure factor $S_D(\pi,\pi)$ (scaled by 0.05) and spin structure factor $S_L(\pi,\pi)$ as functions of the nearest-neighbor interaction V at fixed $U=4$ for (a) $N_h=2$, (b) $N_h=3$, and (c) $N_h=4$ hole-doped systems.}
\label{fig:234hcssfu4}
\end{figure}

Figure \ref{fig:234hcssfu4} shows the evolution of the charge and spin structure factors $S_D(\pi,\pi)$ and $S_L(\pi,\pi)$ as functions of the nearest-neighbour interaction $V$ at fixed $U=4$ for hole dopings $N_h=2,3,4$. The half-filled ($N_h=0$) and one-hole ($N_h=1$) doped systems (results not shown) display very similar trends: $S_L(\pi,\pi)$ is maximal near $V=0$ ($\approx 0.72$) and falls to $\approx 0.19$ by $V=-4$, remaining roughly saturated near 0.2 down to $V=-10$; on the repulsive side the system crosses over from SDW to CDW behaviour near $V\approx U/2$ and CDW correlations dominate for $V\gtrsim 2$.
For $N_h=2$ (Fig. {\color{blue} 9(a)}) the spin structure factor peaks modestly ($\approx 0.35$) at $V\approx -2$ and then decreases to $\approx 0.08$ by $V=-6$, saturating for stronger attraction; as $V$ is increased past the crossover region ($V\sim 2$) charge correlations grow and by $V\gtrsim 4$ $S_D(\pi,\pi)$ dominates, reaching a plateau near 0.40 up to $V=10$. The $N_h=3$ system (Fig. {\color{blue} 9(b)}) follow a similar pattern, except that the negative $V$ saturation of $S_L(\pi,\pi)$ occurs at a much smaller value ($\approx 0.025$ for $V\leqslant -6$). For $N_h=4$ (Fig. {\color{blue} 9(c)}) the spin response peaks around $V=-4$ ($\approx 0.32$) and then collapses to essentially zero at $V=-6$ and beyond; on the repulsive side the same CDW growth seen for $N_h=2$ appears for $V\gtrsim 4$. Across all dopings the charge structure factor is essentially suppressed in the negative $V$ regime and only becomes large on the repulsive side after the SDW$\rightarrow$CDW crossover. 

These results align closely with the binding-energy trends: in the attractive $V<0$ regime, all binding energies ($E_{B2}$,$E_{B3}$ and $E_{B4}$) are negative, and their magnitudes increase with $|V|$, indicating a transition toward phase-separated states with suppressed long-range order---consistent with the nearly vanishing $S_D(\pi,\pi)$ and diminished $S_L(\pi,\pi)$.
In the repulsive $V>0$ regime, both $E_{B2}$ and $E_{B3}$ remain negative while $E_{B4}$ becomes positive, signaling that pair and three-hole binding coexist with enhanced CDW correlations, whereas complete multi-hole clustering is inhibited. The crossover near $V\approx U/2$ thus marks a transition from attraction-driven local pairing to repulsion-stabilized charge ordering where bound holes begin to correlate with the emerging CDW background.

\begin{figure}[h]
\centering
\includegraphics[scale=0.25]{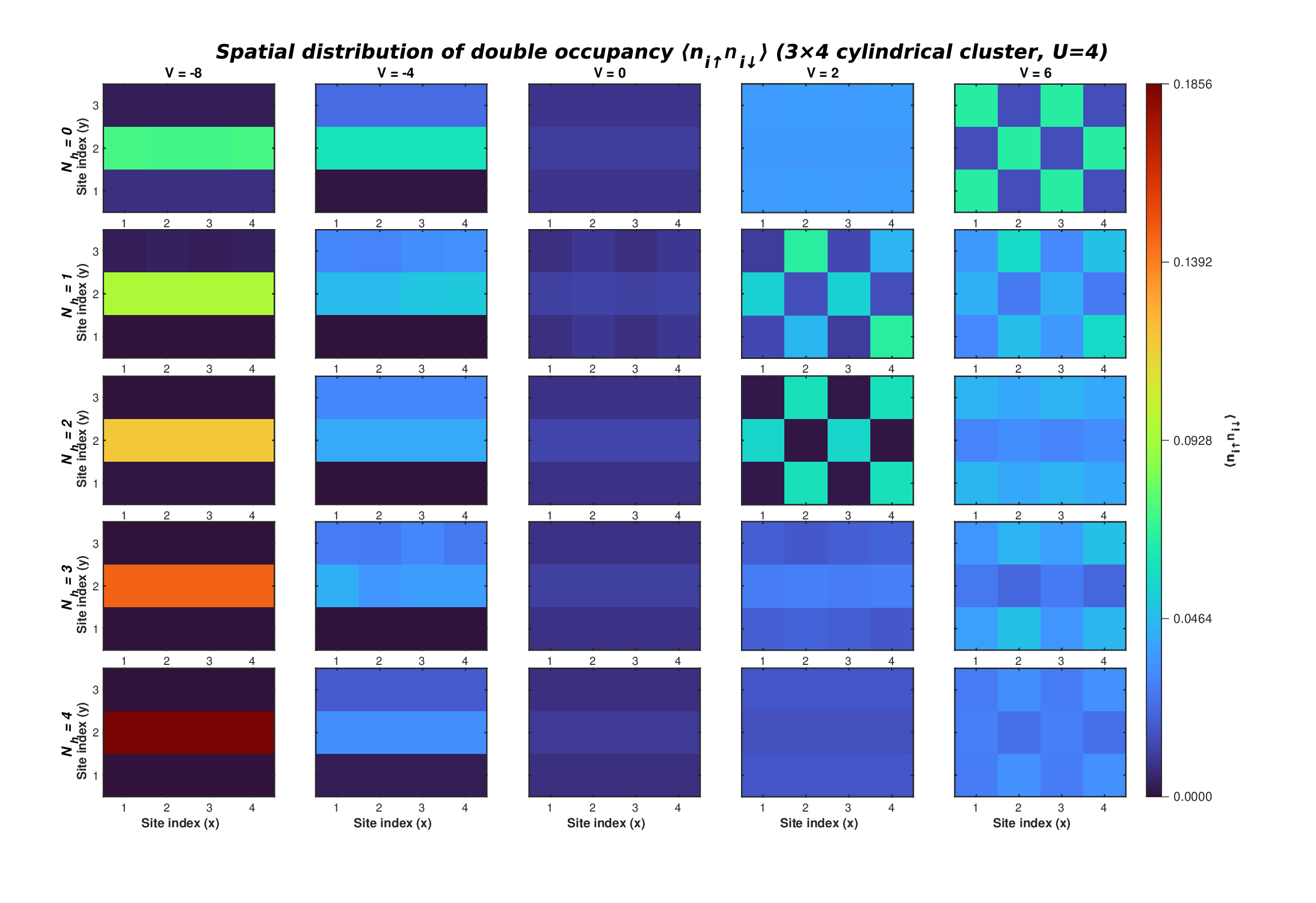}
\caption{Site-resolved double occupancy $\langle n_{i\uparrow}n_{i\downarrow}\rangle$ on the $3\times 4$ cylindrical lattice at fixed $U=4$. Columns (left$\rightarrow$right) correspond to nearest-neighbour interactions $V$=-8,-4,0,2,6; rows (top$\rightarrow$bottom) correspond to hole numbers $N_h$=0,1,2,3,4.}
\label{fig:dblcpsu4}
\end{figure} 

Figure \ref{fig:dblcpsu4} displays the site-resolved double occupancy $\langle n_{i\uparrow}n_{i\downarrow}\rangle$ on the $3\times 4$ cylindrical lattice for five representative values of the nearest-neighbour interaction ($V$=-8,-4,0,2,6) and for hole dopings $N_h$=0-4.
Two robust trends are evident. First, strong intersite attraction ($V<0$) produces site-selective enhancement of on-site doublons. For the strong attractive $V$ the middle row of the cylinder shows markedly larger $\langle n_{i\uparrow}n_{i\downarrow}\rangle$ than the other rows, and this enhancement grows with hole number. This behavior indicates that attractive $V$ drives localized charge clustering at preferred lattice positions selected by the finite-size geometry --- a real-space signature consistent with the large negative multi-hole binding energies in the attractive regime.
Second, repulsive or weakly attractive $V$ suppresses on-site doublons and produces more spatial modulation consistent with incipient CDW order. For $V=2$ and $V=6$, we observe an overall reduction in $\langle n_{i\uparrow}n_{i\downarrow}\rangle$, but a checkerboard-like modulation appears in some dopings, matching the growth of the charge structure factor $S_D(\pi,\pi)$ in the repulsive regime. Taken together with the binding-energy and structure-factor results, the double-occupancy maps show that (i) attraction-driven binding is realized through local on-site doublon enhancement and site-selective clustering, while (ii) repulsion-driven binding is associated with reduced on-site doublons and with spatial charge modulation that favors intersite correlations rather than strong local doublons.

\begin{figure}[h]
\centering
\includegraphics[scale=0.38]{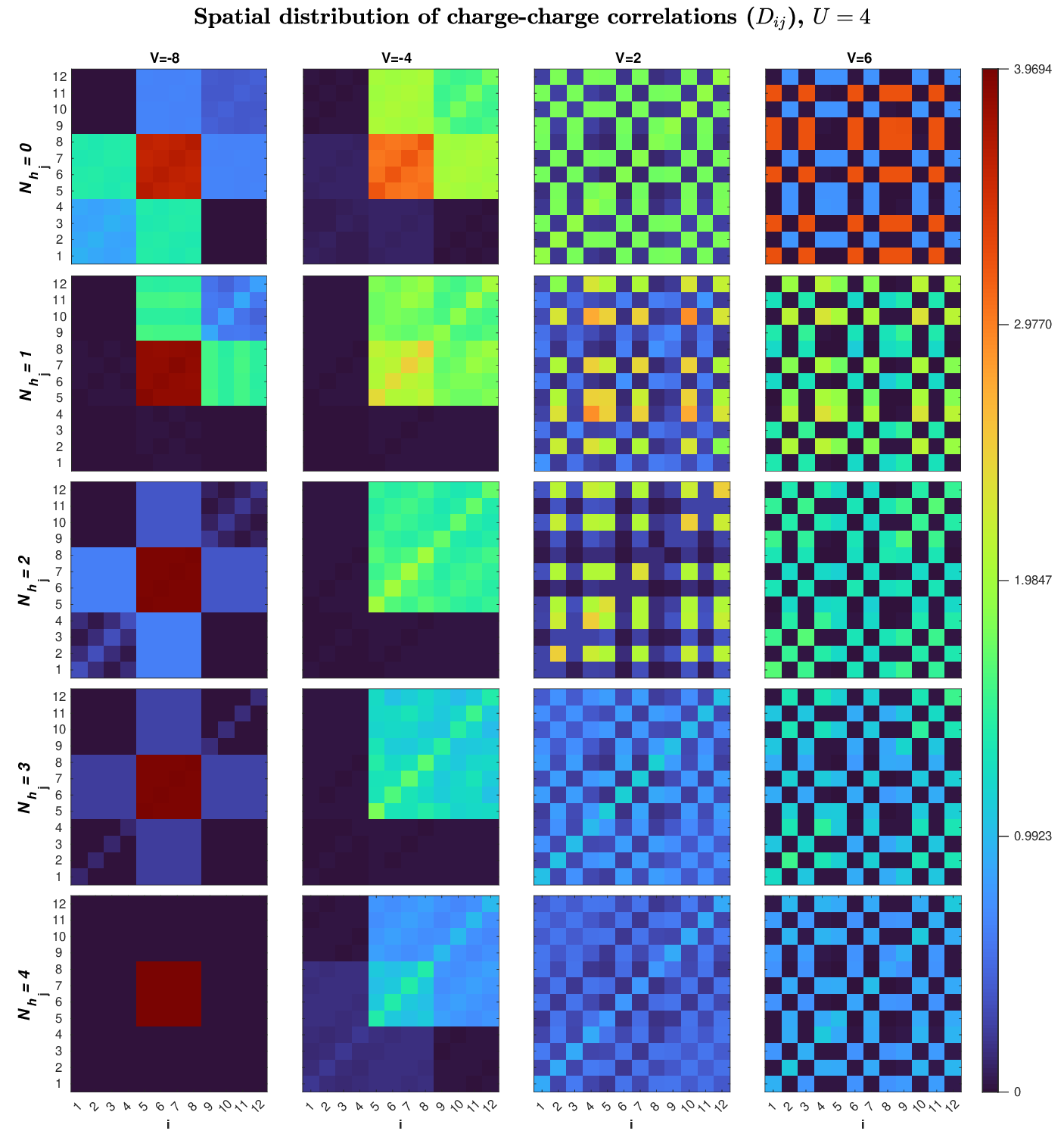}
\caption{Spatial distribution of charge-charge correlations $D_{ij}$ on the $3\times 4$ cylindrical lattice at fixed $U=4$. Each row corresponds to a fixed hole doping $N_h=0-4$; each column corresponds to a representative nearest-neighbor interaction $V$=-8,-4,2,6.}
\label{fig:chargecorru4}
\end{figure} 

To capture the spatial organization of charge fluctuations beyond global structure-factor information, Fig. \ref{fig:chargecorru4} displays the full site-resolved charge-charge correlation matrices for hole dopings $N_h=0,1,2,3,4$ at fixed $U=4$ and representative nearest-neighbor interactions $V$=-8,-4,2,6.
Across the full interaction range, the site-resolved charge-charge correlation matrices $D_{ij}$ reveal a clear and physically consistent reorganization of charge fluctuations driven by the nearest-neighbor interaction $V$. In the strongly attractive regime ($V=-8$), charge fluctuations collapse into a sharply localized region, visible as an intense, compact block of large positive correlations. This localization persists for all dopings, though its spatial extent gradually shrinks with increasing hole number. The absence of any extended modulation and the dominance of a single localized cluster mirror the large negative binding energies $E_{B2}$, $E_{B3}$, and $E_{B4}$ observed for $V\leqslant -2$, confirming that attractive intersite interactions favor real-space clustering rather than long-wavelength ordering. As the attraction weakens ($V=-4$), this isolated charge cluster broadens into a smoother, diagonally extended correlation plateau, signaling that attractive interactions still correlate charge fluctuations within a contiguous region of the lattice but no longer enforce extreme localization. Upon crossing into the weakly repulsive regime ($V=2$), the correlation matrices undergo a qualitative transformation: an alternating, checkerboard-like structure emerges across the entire lattice, indicating the onset of commensurate charge-density-wave (CDW) correlations consistent with the growth of $S_D(\pi,\pi)$ beyond the SDW$\rightarrow$CDW crossover at $V\approx U/2$. This periodic modulation becomes fully developed at strong repulsion ($V=6$), producing a robust, high-contrast bipartite pattern that persists for all dopings despite a gradual reduction in correlation amplitude. This real-space ordering aligns with the suppression of four-hole binding (positive $E_{B4}$) for $V>0$ and the coexistence of residual two- and three-hole binding with the emerging CDW background. Altogether, the charge-correlation patterns establish a coherent picture: attractive $V$ induces localized clustering that drives multi-hole binding and phase-separation tendencies, whereas repulsive $V$ stabilizes extended CDW correlations that suppress large-cluster binding while allowing limited two- or three-hole pairing to survive.

\begin{figure}[h]
\centering
\includegraphics[scale=0.38]{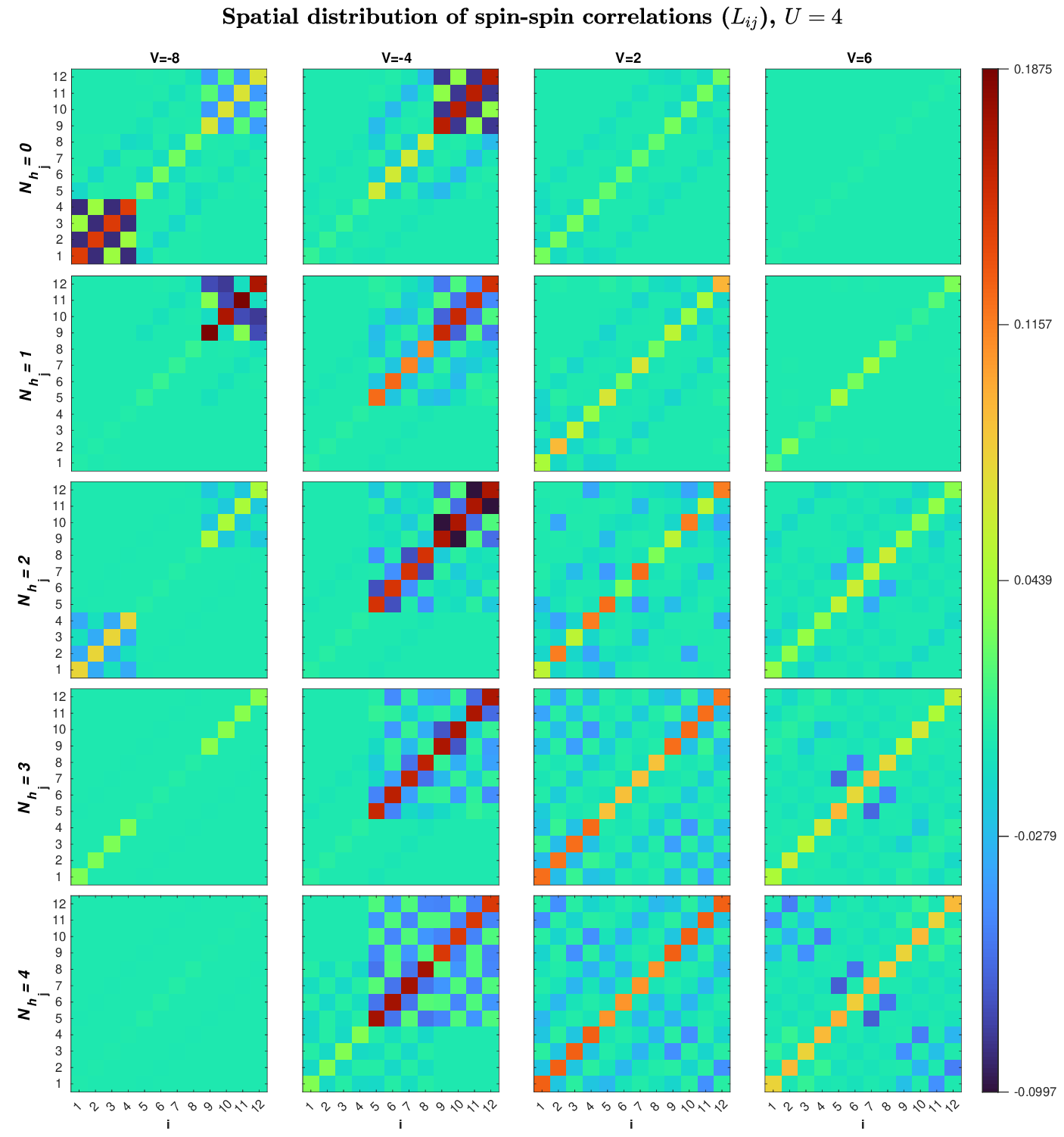}
\caption{Spatial distribution of spin-spin correlations $L_{ij}$ on the $3\times 4$ cylindrical lattice at fixed $U=4$. Each row corresponds to a fixed hole doping $N_h=0-4$; each column corresponds to a representative nearest-neighbor interaction $V$=-8,-4,2,6.}
\label{fig:spincorru4}
\end{figure}

The site-resolved spin-spin correlation matrices $L_{ij}$ for fixed $U=4$ plotted in Fig. \ref{fig:spincorru4}, further clarify how intersite interactions reshape the magnetic landscape across doping and complement the charge-correlation results. 
In the strongly attractive regime ($V=-8$), compact AFM islands emerge in well-defined spatial blocks, reflecting localized spin textures surrounding the charge clusters identified in the corresponding $D_{ij}$ maps shown in Fig. \ref{fig:chargecorru4}. As holes concentrate within a small real-space region, magnetic correlations become increasingly restricted to the remaining singly-occupied sites, producing intense, localized AFM patches for $N_h=0-2$ that rapidly diminish for higher dopings, in agreement with the collapse of $S_L(\pi,\pi)$ at large negative $V$. At moderate attraction ($V=-4$), the AFM pattern extends over larger regions with a clear sublattice alternation, consistent with a more delocalized charge distribution and with the modest enhancement of the spin structure factor near $V\approx -2$. By contrast, in the repulsive regime ($V=2$), the spin correlations become predominantly short-ranged and nearly uniform, indicating the suppression of long-range AFM order as charge modulation begins to dominate. At strong repulsion ($V=6$), the spin maps are almost featureless apart from faint checkerboard oscillations, reflecting residual local moments embedded in a robust CDW background. This progressive suppression of extended magnetic correlations with increasing $V$ is fully consistent with the evolution of the binding energies---where $E_{B4}$ becomes positive while $E_{B2}$ and $E_{B3}$ remain weakly bound---and reinforces the physical picture that attractive $V$ drives localized AFM textures around clustered holes, whereas repulsive $V$ stabilizes charge ordering and quenches spin coherence.

Taken together, the binding energies, structure factors, and site-resolved charge and spin correlation maps present a coherent physical picture for the extended Hubbard model at $U=4$. The attractive intersite interactions ($V<0$) results in strong real-space hole clustering and enhance on-site doublon formation, leading to simultaneously negative $E_{B2}$, $E_{B3}$, and $E_{B4}$ for $V\leqslant -2$, and producing localized magnetic textures confined to the boundaries of charge-rich regions. In contrast, repulsive intersite interactions ($V>0$) stabilize checkerboard-like CDW correlations that suppress multi-hole clustering: $E_{B4}$ becomes positive while $E_{B2}$ and $E_{B3}$ remain weakly negative, and $S_D(\pi,\pi)$ overtakes $S_L(\pi,\pi)$ as the dominant collective response. This interplay between charge order and hole-binding at intermediate coupling highlights the intrinsically competing nature of spin and charge fluctuations in the extended Hubbard model. 

\subsection{Extended Hubbard model: $U=10$}

Having established the microscopic mechanisms governing hole-binding and spin-charge organization at $U=4$, we now turn to the strong-coupling regime $U=10$, where on-site repulsion dominates the energy landscape. In this regime, the doublon sector is strongly suppressed, and the competition between intersite interactions and magnetic background takes on a qualitatively different character. The following section examines how the physics of hole-binding and collective ordering evolves when charge fluctuations are heavily constrained by large on-site repulsion.

\begin{figure}[h]
\centering
\includegraphics[scale=0.64]{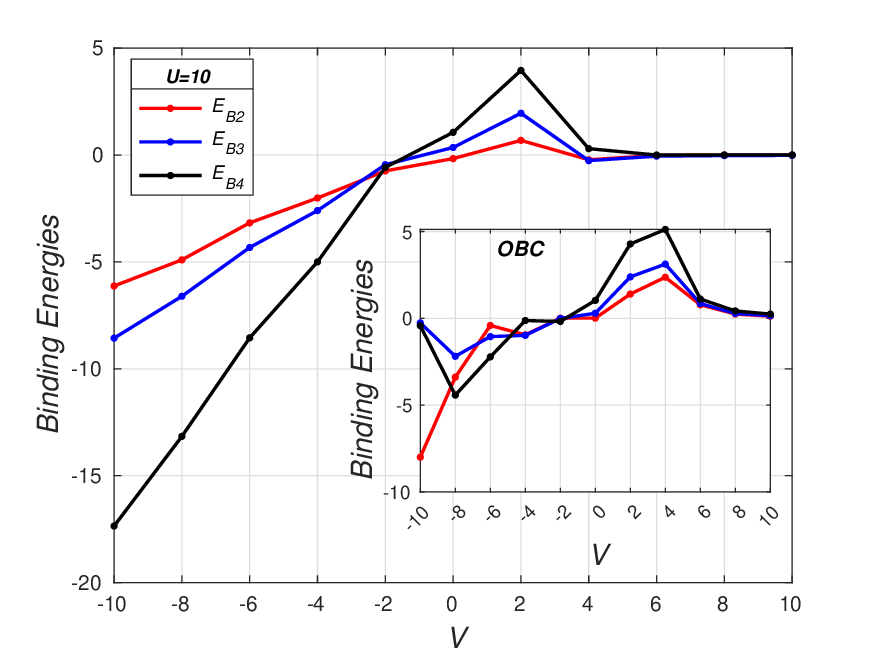}
\caption{Two-hole ($E_{B2}$), three-hole ($E_{B3}$) and four-hole ($E_{B4}$) binding energies as a function of nearest-neighbor interaction $V$ for the extended Hubbard model at $U=10$ on $3\times 4$ cylindrical lattice. The inset compares the corresponding results for the lattice with open boundary conditions.}
\label{fig:be234u10}
\end{figure}

Figure \ref{fig:be234u10} shows the evolution of the two-, three- and four-hole binding energies in the extended Hubbard model at strong coupling $U=10$ as the nearest-neighbour interaction $V$ is varied on the $3\times 4$ cylindrical lattice, with the inset showing the corresponding binding energies for lattice with open boundary conditions. Three clear regimes emerge. For strong attraction ($V\leqslant -2$), all binding energies are substantially negative and their magnitudes grow rapidly with $|V|$, signalling a robust collapse of the doped holes into multi-hole clusters. This is the natural strong-coupling analogue of the $U=4$ behaviour, where moderately attractive $V$ was already sufficient to induce simultaneous negativity of $E_{B2}$, $E_{B3}$ and $E_{B4}$. Here, however, because double occupancy is strongly suppressed by the large $U$, substantially stronger intersite attraction is required to overcome local repulsion and trigger phase separation. Near the weak-interaction window around $V\approx 0$, the system reverts to the behaviour of the pure Hubbard model: only the two-hole binding energy $E_{B2}$ remains weakly negative, while $E_{B3}$ and $E_{B4}$ are positive. This indicates that, in contrast to the $U=4$ case, higher-order clustering is strongly suppressed at $U=10$ when intersite interactions are weak, and the only remnant bound object is a dilute two-hole pair. Moving into the repulsive $V$ regime, the cylindrical lattice exhibits a subtle re-emergence of small negative $E_{B2}$ and $E_{B3}$ for $V\geqslant 4$, while $E_{B4}$ remains positive. Thus, at large $U$ the system supports local two- and three-hole bound states embedded in a repulsion-stabilized charge-modulated background, but no macroscopic phase separation. These local bound objects are absent in the lattice with OBC, where all three binding energies remain positive for $V\geqslant 0$, demonstrating that finite-size geometry and boundary conditions can stabilize nontrivial pairing tendencies even in a globally repulsive environment. At moderately negative $V$ ($V=-2$), in the OBC lattice the onset of multi-hole clustering occurs slightly earlier ($E_{B3}$, $E_{B4}<0$ while $E_{B2}>0$), again reflecting boundary-induced differences in kinetic-energy optimization on small lattices. Taken together, the $U=10$ binding-energy systematics show that strong on-site repulsion substantially sharpens the phase-separation threshold, eliminates higher-order clustering in the weak-interaction regime, and enhances the sensitivity of the repulsive $V$ sector to geometry---features that stand in sharp contrast to the richer and more symmetric hole-binding landscape observed at $U=4$.

\begin{figure}[h]
\centering
\includegraphics[scale=0.25]{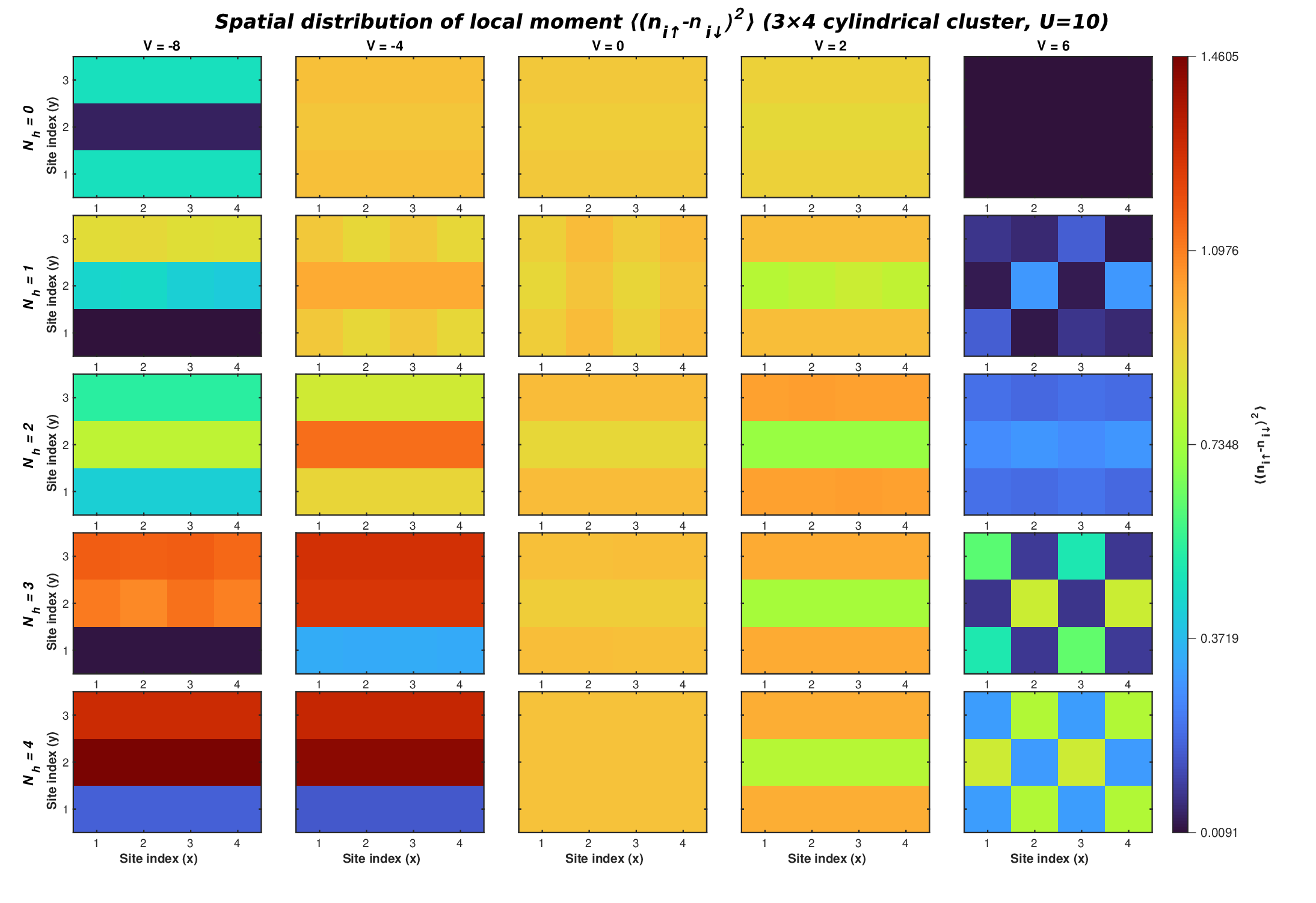}
\caption{Site-resolved local moment $\langle (n_{i\uparrow}-n_{i\downarrow})^2\rangle$ on the $3\times 4$ cylindrical lattice at fixed $U=10$. Columns (left$\rightarrow$right) correspond to nearest-neighbour interactions $V$=-8,-4,0,2,6; rows (top$\rightarrow$bottom) correspond to hole numbers $N_h$=0,1,2,3,4.}
\label{fig:lclmu10}
\end{figure} 

To complement the binding energy analysis and elucidate the magnetic landscape in the strong-coupling regime, we examine the site-resolved local moment $\langle (n_{i\uparrow}-n_{i\downarrow})^2 \rangle$ on the $3\times 4$ cylindrical lattice for the extended Hubbard model at $U=10$. Figure \ref{fig:lclmu10} displays the spatial distribution across hole dopings $N_h=0-4$ and representative nearest-neighbour interactions $V=-8,-4,0,2,6$. 
At $V=0$, corresponding to the simple Hubbard model at strong coupling, the local moment reveals characteristic doping-dependent behavior. For the half-filled system ($N_h=0$), moments are uniformly enhanced ($\sim 0.92-0.94$) across the lattice, reflecting the Mott insulating state where strong on-site repulsion suppresses double occupancy and stabilizes local magnetic moments. With one-hole doping ($N_h=1$), moments remain robust but develop spatial modulation, with certain sites (particularly in the middle row) showing enhanced values ($\sim 0.97$) while others are slightly reduced ($\sim 0.89-0.92$), indicating the formation of a magnetic polaron where the hole locally disrupts the antiferromagnetic background. The two-hole doped system ($N_h=2$) exhibits a similar pattern with pronounced moment enhancement ($\sim 0.97$) on specific sites, consistent with the formation of a bound pair that locally perturbs but does not destroy the magnetic environment. This magnetic texture supports the weakly negative $E_{B2}$ observed at $V=0$. For higher dopings ($N_h$=3,4), moments become more uniformly distributed with moderate suppression ($\sim 0.95-0.96$), reflecting the disrupted magnetic coherence and consistent with the positive three- and four-hole binding energies.
The introduction of attractive intersite interactions ($V<0$) dramatically reshapes the magnetic landscape, driving a profound real-space reorganization. In the strongly attractive regime ($V=-8$), the local moments collapse into sharply defined spatial domains. For $N_h=1$, the entire bottom row exhibits near-zero moment ($\sim 0.010$), indicating complete magnetic quenching and the formation of a coherent, charge-rich region. The adjacent middle row shows moderately suppressed moments ($\sim 0.41-0.45$), while the top row retains relatively enhanced moments ($\sim 0.86-0.89$). This pattern intensifies with doping: at $N_h=2$, the bottom row remains magnetically quenched ($\sim 0.42$), the middle row shows intermediate moments ($\sim 0.79$), and the top row is further suppressed ($\sim 0.54$). For $N_h=3-4$, the entire lattice shows strongly suppressed moments, with $N_h=4$ exhibiting a striking pattern where the bottom row is quenched ($\sim 0.17$), while the middle and top rows show anomalously high moments ($\sim 1.46$ and $\sim 1.29$, respectively), suggesting a phase-separated state with hole-rich and hole-poor magnetic domains. This real-space magnetic collapse directly mirrors the large negative binding energies $E_{B2}$, $E_{B3}$, and $E_{B4}$ observed for $V\leqslant -2$, confirming that attractive $V$ drives phase separation into magnetically quenched, hole-rich clusters and moment-preserving, hole-poor regions.
In the repulsive $V$ regime ($V>0$), the local moment distribution reveals a more subtle interplay between charge order and magnetic correlations. At $V=2$, moments become more uniform than in the $V=0$ case, with a slight overall suppression and the emergence of weak spatial modulation. This reflects the competition between on-site repulsion and intersite repulsion, which suppresses hole clustering. At $V=6$, a clear bipartite pattern emerges across all dopings, with moments alternating between higher ($\sim 0.76-0.83$) and lower ($\sim 0.30-0.31$) values on neighboring sites in a checkerboard arrangement. This modulation aligns with the charge-density-wave (CDW) correlations identified in the charge structure factor $S_D(\pi,\pi)$ (results not shown) and represents the magnetic signature of the underlying CDW order: sites with enhanced electron density exhibit stronger local moments, while hole-rich sites show moment suppression. The persistence of this pattern even in the presence of two- and three-hole binding (negative $E_{B2}$ and $E_{B3}$ at $V=6$) indicates that bound holes in the repulsive regime form within, and are constrained by, the emerging charge-ordered background, rather than collapsing into compact clusters as in the attractive case.

The evolution of the local moment across interaction regimes provides crucial microscopic insight into the magnetic underpinnings of hole-binding. In the attractive $V$ regime, the formation of extended magnetically quenched domains confirms that hole clustering effectively destroys local magnetic order within cluster regions. In contrast, repulsive $V$ preserves a modulated moment pattern that coexists with bound pairs, demonstrating that hole-binding in this regime occurs within a magnetically active, charge-ordered environment rather than through phase separation. These distinct magnetic signatures reinforce the physical picture drawn from binding energies analyses: attractive interactions drive macroscopic phase separation with magnetic domain formation, while repulsive interactions stabilize bound holes within a correlated CDW background that maintains finite local moments.

\begin{figure}[h]
\centering
\includegraphics[scale=0.38]{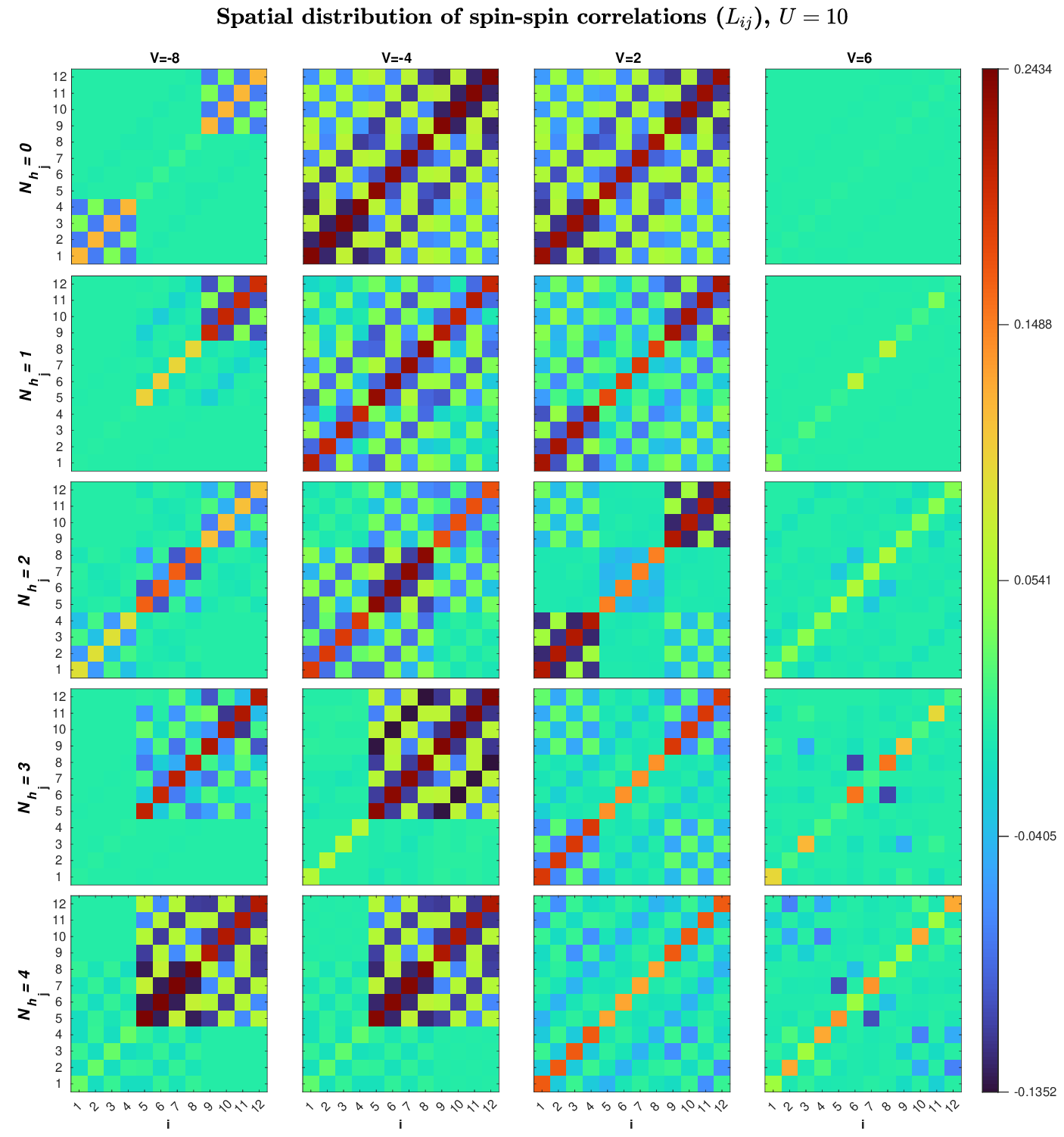}
\caption{Spatial distribution of spin-spin correlations $L_{ij}$ on the $3\times 4$ cylindrical lattice at fixed $U=10$. Each row corresponds to a fixed hole doping $N_h=0-4$; each column corresponds to a representative nearest-neighbor interaction $V$=-8,-4,2,6.}
\label{fig:spincorru10}
\end{figure} 

The spatial distributions of the spin-spin correlations $L_{ij}$ at $U=10$ (Fig. \ref{fig:spincorru10}) provide a real-space complement to the local-moment analysis and reveal how the intersite interaction $V$ governs magnetic textures in the strong-coupling regime. Under strong attraction ($V=-8$), the correlation matrices exhibit compact blocks of intense correlations surrounded by nearly inert regions, indicating phase-separated clusters where magnetic moments are quenched within hole-rich domains while antiferromagnetic (AFM) order persists locally in the hole-poor background. At moderate attraction ($V=-4$), the correlations extend across the lattice in a clear checkerboard-like alternation, showing that partial clustering still allows sizeable AFM coherence on singly occupied sites. For weakly repulsive $V=2$, the maps display strong diagonal elements and a relatively uniform short-range AFM pattern, consistent with a magnetically active but spatially constrained background that hosts the weak two-hole binding observed near $V\approx 0$. In contrast, strong repulsion ($V=6$) suppresses long-range spin coherence, leaving only faint checkerboard remnants that coincide with the charge-density-wave (CDW) modulation evident in the local-moment maps. Across increasing hole number, localized distortions of the AFM background---magnetic-polaron-like features---appear for one and two holes, while higher doping leads to global magnetic suppression at attractive $V$ and CDW-constrained spin correlations at repulsive $V$. Overall, the $L_{ij}$ patterns confirm that attractive interactions induce hole clustering accompanied by magnetic quenching, whereas repulsive $V$ confines spin correlations to electron-rich sublattices, completing a consistent picture of charge-spin interplay established by the binding-energy and local-moment analyses.

\begin{figure}[h]
\centering
\includegraphics[scale=0.38]{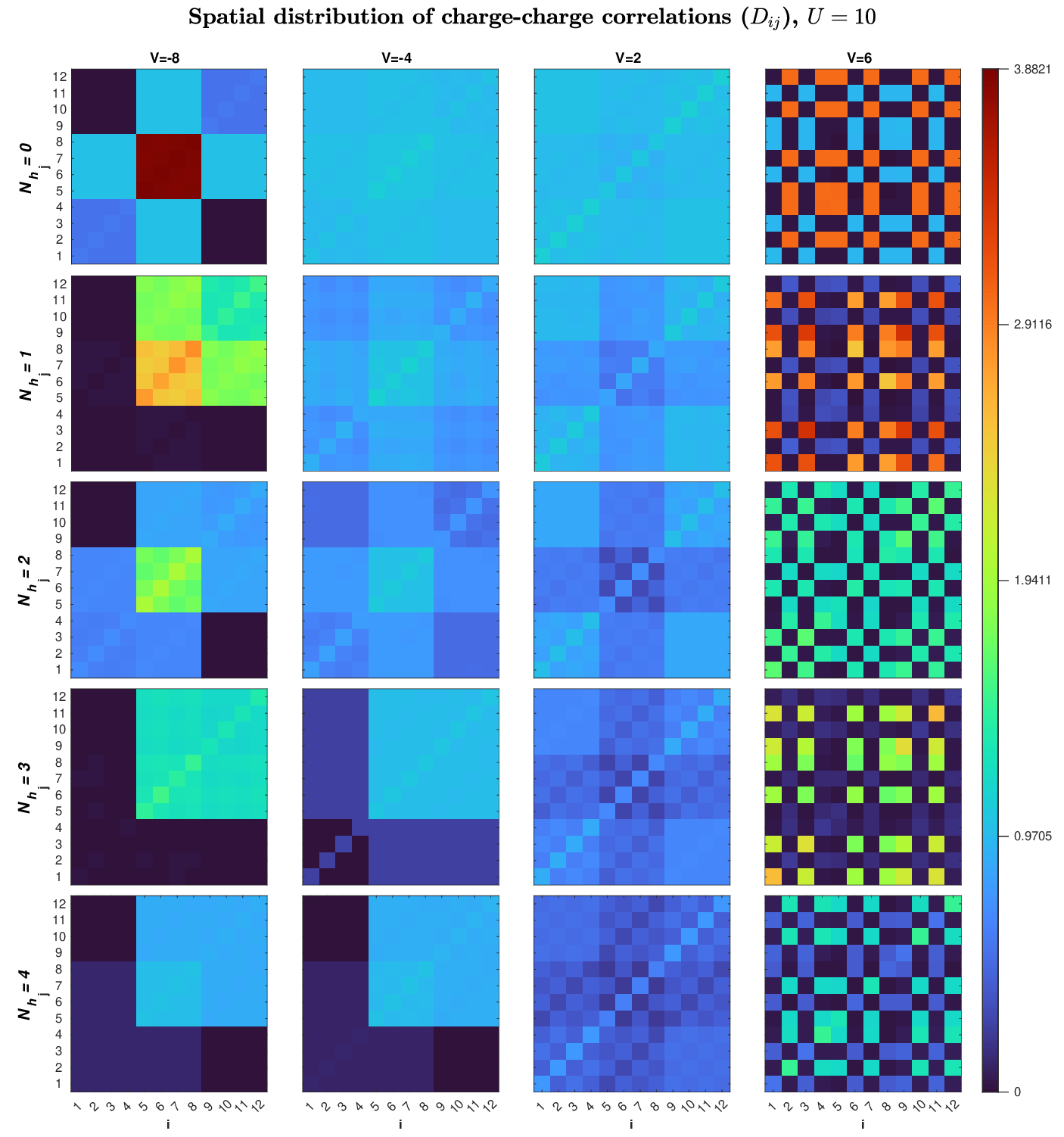}
\caption{Spatial distribution of charge-charge correlations $D_{ij}$ on the $3\times 4$ cylindrical lattice at fixed $U=10$. Each row corresponds to a fixed hole doping $N_h=0-4$; each column corresponds to a representative nearest-neighbor interaction $V$=-8,-4,2,6.}
\label{fig:chargecorru10}
\end{figure} 

The site-resolved charge-charge correlation matrices $D_{ij}$ at $U=10$ (Fig. \ref{fig:chargecorru10}) further illuminate the real-space reorganization of charge induced by nearest-neighbour interaction $V$ and tie directly to the binding-energy and local-moment trends discussed above. Three interaction regimes are immediately apparent. Under strong attraction ($V=-8$) the $D_{ij}$ maps display a compact, high-amplitude block of correlations that is localized to a small region of the lattice for low-to-moderate doping and grows with hole number; this intense localized signal is the charge-space fingerprint of multi-hole clustering and phase separation and matches the magnetic quenching and strongly negative multi-hole binding energies observed in this regime. At moderate attraction ($V=-4$) the charge correlations are still enhanced relative to the repulsive side but are more spatially distributed: the correlation block broadens into a contiguous but less concentrated plateau, signalling partial clustering where charge is correlated over a region rather than collapsed onto a single site cluster---consistent with the more extended AFM texture and intermediate doublon suppression seen at $V=-4$. For weak-to-moderate repulsion ($V=2$) the matrices become fairly uniform with only modest, local modulations; charge correlations are weaker and more homogeneous, reflecting the suppression of strong on-site doublons by large $U$ and the survival of only short-range charge coherence that can host weak two-hole binding near $V\approx 0$. Finally, in the strongly repulsive regime ($V=6$) a pronounced checkerboard pattern appears across the entire lattice for all dopings: high correlations on one sublattice and low correlations on the other indicate the emergence of commensurate CDW order stabilized by intersite repulsion. The amplitude and sharpness of this bipartite pattern vary with doping, but its presence explains the suppression of four-hole clustering and the persistence of residual two- and three-hole binding embedded in a CDW background. Across dopings the one- and two-hole cases show clear signatures of localized charge distortions that are progressively replaced by global clustering at strong negative $V$ or by sublattice-separated CDW order at large positive $V$. Collectively, the $D_{ij}$ maps at $U=10$ provide a consistent real-space picture: attractive $V$ promotes localized, phase-separated charge clusters (and concomitant magnetic collapse), whereas repulsive $V$ drives extended, bipartite charge ordering that constrains but does not entirely eliminate locally bound hole-pairs.

Overall, the extended Hubbard model at strong coupling ($U=10$) exhibits a clear, interaction-driven dichotomy: sufficiently attractive nearest-neighbour interaction $V$ overcome the large on-site repulsion $U$ to produce robust multi-hole clustering, magnetic quenching in cluster regions, and large negative binding energies, while repulsive intersite interactions stabilize checkerboard CDW order, suppress macroscopic phase separation, and confine magnetic correlations to the electron-rich sublattice. Near $V\approx 0$ the system retains a vestigial tendency for weak two-hole binding embedded in a magnetically active background, but higher-order clustering is strongly suppressed by the large $U$. Finite-size geometry and boundary conditions modestly affect the precise thresholds and amplitudes; nevertheless, the $U=10$ data robustly demonstrate that nonlocal interactions qualitatively reshape whether hole binding proceeds via phase separation (attraction) or via pairing inside a charge-ordered background (repulsion).
\section{\label{summary}Summary and Conclusions}
We presented an exact diagonalization study on the $3\times 4$ cylindrical lattice for the simple and extended Hubbard model, revealing a rich landscape of hole-binding and competing spin-charge order in $U,V$ parameter space. The analysis demonstrates that the competition between the on-site repulsion $U$ and the nearest-neighbor interaction $V$ leads to distinct phases characterized by different pairing mechanisms and magnetic environments.

In the simple Hubbard model ($V=0$), we identified a crossover near intermediate values of $U$ where bound-hole states coexist with both spin- and charge-density modulations. The two-hole binding energy $E_{B2}$ remains weakly negative for repulsive $U$, indicating correlation-mediated pairing, while the three- and four-hole binding energies remain positive, thereby ruling out the phase separation. This pairing arises within a magnetically correlated environment where robust nearest-neighbor antiferromagnetic correlations ($L_{1,2}$) coexist with sign-reversed next-nearest-neighbor correlations ($L_{1,3}$), creating a magnetic texture that facilitates hole motion.

The extended Hubbard model at intermediate coupling ($U=4$) exhibits a dramatic dependence on intersite interactions. Attractive $V$ drives real-space hole clustering, evidenced by simultaneously negative $E_{B2}$, $E_{B3}$, and $E_{B4}$ for $V\leqslant -2$, accompanied by localized doublon enhancement and fragmented antiferromagnetic islands. In contrast, repulsive $V$ stabilizes checkerboard charge-density-wave (CDW) order that suppresses multi-hole clustering while allowing two- and three-hole binding to persist within the CDW background. The SDW-to-CDW crossover near $V\approx U/2$ marks a fundamental transition in the system's dominant ordering tendency.

At strong coupling ($U=10$), the physics of hole-binding becomes more sharply defined. The attractive $V$ overcomes the large on-site repulsion $U$ to induce phase separation with magnetically quenched hole-rich clusters, while repulsive $V$ produces a robust CDW state where bound holes exist within the charge-ordered background. The local moment and spin correlation maps reveal that attractive interactions destroys magnetic coherence in clustered regions, whereas repulsive interactions preserve modulated moments constrained by the CDW ordering.

These findings establish a coherent microscopic picture of hole-binding in correlated systems: attraction-driven phase separation proceeds through magnetic domain formation and charge collapse, while repulsion-mediated binding occurs within magnetically active, charge-ordered environments. The distinct real-space signatures of these mechanisms---localized clustering versus extended CDW modulation---provide clear criteria for identifying pairing regimes in finite-cluster studies.

Building on these findings, several natural extensions would strengthen the conclusions and connect them more directly to experiments. Larger-system and quasi-2D studies using complementary methods such as DMRG on wider cylinders, tensor networks and cluster DMFT would test the persistence of the observed cluster or paired regimes in the thermodynamic limit. Finite-temperature calculations and evaluation of dynamical response functions would clarify whether the binding tendencies identified here produce low-energy pairing signatures likely to survive fluctuations. These avenues will determine whether the local binding and competing spin-charge textures documented on small clusters seed genuine superconducting correlations in extended systems or instead signal the proximity to phase separation.

In summary, our ED results on the $3\times 4$ cylindrical lattice demonstrate that nonlocal interactions qualitatively control whether doped holes bind via phase separation (attractive $V$) or form local bound states inside a charge-ordered background (repulsive $V$), with the on-site repulsion $U$ setting the doublon cost and sharpening the associated crossovers. These microscopic insights clarify how spin and charge channels cooperate and compete to produce distinct pairing and ordering tendencies in minimal correlated-electron models.
\section*{Acknowledgments}

MFE and MAHA acknowledge the support from National PARAM Supercomputing Facility (NPSF), C-DAC, Pune, India, for providing computing facility for part of this work.


\bibliography{main}

\end{document}